\pdfoutput=1

\documentclass[%
 reprint,
 superscriptaddress,
 nobibnotes,
 amsmath,amssymb,
 aps,
 showkeys,
 longbibliography,
]{revtex4-2}

\usepackage{graphicx}% Include figure files
\usepackage{dcolumn}% Align table columns on decimal point
\usepackage{bm}% Bold math
\usepackage{chemformula}% Chemical formulas
\usepackage[separate-uncertainty=true,separate-uncertainty,multi-part-units=single]{siunitx}% SI units with separate uncertainty
\usepackage{hyperref}% Add hypertext capabilities 
\hypersetup{colorlinks=true,allcolors=magenta}

\begin{document}

\title{Alcohol-Based Adsorption Heat Pumps using Hydrophobic Metal-Organic Frameworks}

\author{R. M. Madero-Castro}
    \affiliation{Department of Physical, Chemical, and Natural Systems, Universidad Pablo de Olavide. Ctra. Utrera km. 1. ES-41013 Seville, Spain.}
\author{A. Luna-Triguero}
    \affiliation{Department of Mechanical Engineering, Eindhoven University of Technology, 5600 MB Eindhoven, The Netherlands.}
    \affiliation{Eindhoven Institute for Renewable Energy Systems (EIRES), Eindhoven University of Technology, Eindhoven 5600 MB, The Netherlands.}
    \author{C. González-Galán}
    \affiliation{Department of Physical, Chemical, and Natural Systems, Universidad Pablo de Olavide. Ctra. Utrera km. 1. ES-41013 Seville, Spain.}
\author{J. M. Vicent-Luna}
    \email[Corresponding author: ]{j.vicent.luna@tue.nl}
    \affiliation{Materials Simulation \& Modelling, Department of Applied Physics, Eindhoven University of Technology, 5600 MB, Eindhoven, The Netherlands.}
    \affiliation{Eindhoven Institute for Renewable Energy Systems (EIRES), Eindhoven University of Technology, Eindhoven 5600 MB, The Netherlands.}
    \author{S. Calero}
    \email[Corresponding author: ]{s.calero@tue.nl}
    \affiliation{Materials Simulation \& Modelling, Department of Applied Physics, Eindhoven University of Technology, 5600 MB, Eindhoven, The Netherlands.}
    \affiliation{Eindhoven Institute for Renewable Energy Systems (EIRES), Eindhoven University of Technology, Eindhoven 5600 MB, The Netherlands.}

\date{\today}

\begin{abstract}

The building climate industry and its influence on energy consumption have consequences on the environment due to the emission of greenhouse gasses. Improving the efficiency of this sector is essential to reduce the effect on climate change. In recent years, the interest in porous materials in applications such as heat pumps has increased for their promising potential. To assess the performance of adsorption heat pumps and cooling systems, here we discuss a multistep approach based on the processing of adsorption data combined with a thermodynamic model. The process provides properties of interest, such as the coefficient of performance, the working capacity, the specific heat or cooling effect, or the released heat upon adsorption and desorption cycles, and it also has the advantage of identifying the optimal conditions for each adsorbent-fluid pair. To test this method, we select several metal-organic frameworks that differ in topology, chemical composition, and pore size, which we validate with available experiments. Adsorption equilibrium curves were calculated using molecular simulations to describe the adsorption mechanisms of methanol and ethanol as working fluids in the selected adsorbents. Then, using a thermodynamic model we calculate the energetic properties combined with iterative algorithms that simultaneously vary all the required working conditions. We discuss the strong influence of operating temperatures on the performance of heat pump devices. Our findings point to the highly hydrophobic metal azolate framework MAF-6 as a very good candidate for heating and cooling applications for its high working capacity and excellent energy efficiency.

\end{abstract}

\keywords{Refrigeration, Optimal conditions, Energy storage, Molecular simulations}

\maketitle

%%%MAIN TEXT%%%%%%%%%%%%%%%%%%%%%%%%%%%%%%%%%%%%%%%%%%%%
%%%%%%%%%%%%%%%%%%%%%%%%%%%%%%%%%%%%%%%%%%%%%%%%%%%%%%%%
%%%%%%%%%%%%%%%%%%%%%%%%%%%%%%%%%%%%%%%%%%%%%%%%%%%%%%%%
%%%%%%%%%%%%%%%%%%%%%%%%%%%%%%%%%%%%%%%%%%%%%%%%%%%%%%%%

\section{Introduction}
\label{sec:intro}
%%%
The building energy demand for energy efficiency has increased in the last two decades.\cite{Sayadi2019,Perez-Lombard2008} Forecasts manifest that this growth will continue rising due to global warming.\cite{Perez-Lombard2008,Isaac2009} Consequently, the emission of greenhouse gasses into the atmosphere will be higher, producing a feedback process.\cite{Hu2018,Joos2015,Govindasamy2005} Improving efficiency in this sector and reducing  greenhouse gasses emissions are critical aspects in mitigating the climate change.\cite{Hay2015, Liu2019,Coumou2013, Elsheikh2018} Traditional heating systems such as electric or gas heaters offer low performance.\cite{Proszak2020} They are based on Joule effect or on the specific heat of the substances to be burned, respectively, so that the heat supplied for the heating system is equal to the energy used. It is essential to foster advanced heating devices that take advantage of outside heat to increase performance. For example, heating devices like solar collectors, which use solar energy to warm water or air, are eco-friendly options.\cite{Sharma2009,Kumar2017,Kannan2016}  The main problem of solar systems is that weather acts as a restrictive factor.\cite{Wojdyga2008}

Heat pumps are a promising alternative since they take heat from the surroundings, reducing the total energy consumption of the heating system.\cite{Chua2010} These systems are efficient and a sustainable alternative to conventional methods, increasing the performance and decreasing greenhouse gasses released into the atmosphere. Traditional heat pumps are based on the compression/decompression of a working fluid, with hydrofluorocarbons (like HFC-134a or HFC-125) being the most used at the industrial level.\cite{Kuyper2019,Wongwises2005} The main drawback with these devices is that they use greenhouse gasses \cite{Hu2017} that need to be reduced according to the Montreal Protocol.\cite{Hu2017} In this context adsorption-based Heat Pumps (AHP) and Adsorption Cooling Systems (ACS) using porous materials, can be an alternative. The operation mechanism of the AHP and ACS devices is based on the adsorption and evaporation of a refrigerant. These devices follow the assumption that reversible adsorption (desorption) is an exothermic (endothermic) process.\cite{Teker1999} 

In recent years, AHP and ACS made with porous materials such as activated carbons,\cite{Cacciola1995, Kannan2016, Critoph1989, Wang2003, Kohler2017,Trager2020, Tamainot2001,ElSharkawy2016,Askalany2017,Capri2020} zeolites,\cite{Capri2020,Critoph2005,Freni2016,Calabrese2017,Zheng2015, Freni2015, Atakan2013, Tatlier2021,madero-HT-2023} and metal-organic frameworks (MOFs)\cite{Capri2020,deLange2015,Henninger2017,Kummer2017,Jeremias2014, Steinert2020, Pinheiro0000} have shown promising performance and benefits in cost and versatility. Pal \textit{et al}.\cite{Pal2017} studied the production of highly porous carbons from vegetable waste for heat pump applications. Works reported by de Lange \textit{et al.}\cite{deLange2015a}, Li \textit{et al.}\cite{Li2019}, Erdős \textit{et al.}\cite{Erdos2018}, or Jeremias \textit{et al.}\cite{Jeremias2014a}, among others, studied a variety of MOFs with methanol and ethanol as working fluids for adsorption-driven heat pumps and chillers applications. Kayal \textit{et al.}\cite{Kayal2016} studied the water-AQSOA zeolite working pair to create of adsorption chillers, concluding that these zeolites are also suitable for this purpose. One of the most challenging parts of designing AHPs is the selection of the working pair. It is desirable to work with environmentally friendly fluids with high enthalpy of vaporization. However, the large amount of synthesized porous adsorbents makes it difficult to assess the performance of each working pair experimentally. Computationally, testing several operating conditions for the selection of the optimal range for a given pair is also very expensive. The number of synthetized MOFs is large and continuously growing.\cite{Cambridge} The high versatility in terms of composition and pore size makes them good candidates to operate in different conditions. In-depth knowledge of the adsorbent-fluid interactions makes it possible to choose the most suitable pair for given working conditions. In general, MOFs exhibit higher adsorption capacity than zeolites or other porous materials, a fundamental aspect of increasing the heat transfer between adsorption/desorption cycles. In this work, we have selected ZIF-8, ZIF-71, ZIF-90, MIL-140C, and MAF-6, as adsorbents. Four of these MOFs exhibit large pore size (> 11 Å), which is desired for a stepwise isotherm. The exception is MIL-140C, with pore size lower than 6 Å, which we include for comparison. These MOFs have been successfully tested for the adsorption of methanol and ethanol \cite{deLange2015a, Zhang2013, He2015a}. We use these experimental data to validate our models. 

The selection of the refrigerant for an adsorption energy storage device is as important as the selection of the adsorbent, because the synergy between adsorbent and fluid working pairs is a critical aspect of achieving maximum performance. As alternatives to conventional refrigerants, \cite{Critoph2005,Keinath2017} and water \cite{Luna-Triguero2019, Keinath2017} have been proposed as working fluids. Water is an excellent candidate due to the high enthalpy of vaporization and the zero toxicity to humans, but as a downside, water can affect MOFs stability.\cite{Alvarez2017} Ammonia is also an excellent working fluid with a shallow melting point (-40ºC) and slightly lower enthalpy of vaporization than water. However, ammonia is a toxic compound that should be used carefully. In the search for other working fluids for AHPs and ACSs, light alcohols have become a possible alternative.\cite{deLange2015a,Madero-Castro2022} Highly hydrophobic materials cannot adsorb/desorb water at realistic pressure/temperature conditions.\cite{Sann2018,Calero2015,Voorde2015,He2015} However, these adsorbents can capture and release methanol and ethanol within the range of operating conditions. de Lange \textit{et al.}\cite{deLange2015a}  summarized the main differences between replacing water (the most common working fluid in MOFs) with ethanol and methanol. These are, among others: 1) decrease of the onset pressure of methanol and ethanol compared to water, 2) lower hysteresis loop for large pore materials (below 3.4 nm), 3) lower energy release per cycle, but heat and mass transfer may be improved, and 4) in general, MOFs seem to be more stable upon alcohol adsorption than upon water adsorption. These facts make light alcohols a promising alternative to water and other common refrigerants used in industry. Another important aspect is the reduced global warming potential (GWP) of small alcohols compared to traditional working fluids. Ethanol GWP fluctuates between 0.31-5.55 depending on the production path \cite{Pacheco2019} compared to the 1120-3500 GWP for HFC-based refrigerants, i.e. HFC-134a and HFC-125. Hence, in addition to the previous discussion, the relatively low cost, high heat capacities, and low melting points make methanol and ethanol good candidates as working fluids for AHP/ACS applications.

The large amount of synthesized structures makes the assessment of the performance of each working pair challenging from an experimental point of view. Several studies focused on strategies to analyze the performance of alcohol-adsorbent using computational screenings. Erdős \textit{et al.}\cite{Erdos2018} designed a screening process for methanol and ethanol for about 3000 adsorbents. The most promising structures were selected based on the maximum working capacity obtained in consecutive relative values of pressure defined by the authors. A computational screening and selection based on the coefficient of performance (COP) for cooling applications was reported by Li \textit{et al.}\cite{Li2019} They systematically rejected structures that perform below the imposed limit of 0.8. The significant number of samples made an in-depth study challenging in suitable operating conditions for each system, which led to qualitative-based interpretations. To fasten the selection process, they performed relatively short GCMC simulations (4 $\cdot$ 10$^4$ cycles), only running more cycles for the promising structures. This approach was aimed to obtain high-performance structures. However, short simulations in the first step can also lead to the rejection of viable materials. It has been previously demonstrated that materials with big pores and hydrophobic structures need more MC cycles to reach equilibrium.\cite{Luna-Triguero2019} Following a similar approach, another recent computational screening by the same authors \cite{WeiLi2020} analyzed the performance of COF-ethanol working pairs for heating, cooling, and ice-making applications. As before, they performed short GCMC simulations that could disregard promising candidates. Each adsorbent-fluid pair and its range of operating conditions should be analysed carefully. This is because any small change could imply considerable deviation in the predicted application of heating and cooling systems. In this regard, Xia \textit{et al.}\cite{Xia202011} went a step further by varying the working conditions and the effect on the cooling/heating performance of COF-5/ethanol and several MOF/ethanol working pairs. 

Here we propose a novel approach to assess the coefficient of performance or other thermodynamic quantity, such as the heat released during the adsorption process. We apply this method to investigate MOFs-methanol/ethanol working pairs and rely on adsorption data and the consecutive application of mathematical and thermodynamic models. The advantage of this methodology is that it can be applied to either experimental or computational data sets and is also extensible to other working fluids. Another benefit of the analysis proposed here is that we considered a wide range of operating conditions proving that setting the operating temperatures could lead to a significant loss of information about the performance of the process. In short, an AHP cycle consists of four primary parts: an adsorber containing the adsorbent, a condenser, an evaporator, and an expansion valve. The heat pump operates by driving adsorbate between the adsorber, condenser, and evaporator. The cycle can be divided in two parts. In the first part, the evaporator vaporizes the fluid taking heat from a low-temperature source and releasing heat to an intermediate temperature source (adsorption). In the second part, during the condensation of the fluid, the condenser receives heat from a high-temperature source and releases heat to a second intermediate temperature source (desorption). During the cycle, the evaporator, condenser, desorption, and intermediate temperatures play an important role on the performance of the process. Our approach simultaneously screens all these temperatures providing detailed information on the operation of each MOF-fluid working pair.

With this work, we want to understand the mechanisms that govern the adsorbate-adsorbent pair that enhances the efficiency of the thermodynamic cycle based on the adsorption equilibrium properties of ethanol and methanol in MOFs. We search for optimal conditions for each MOF-alcohol pair, discussing the limitations of setting the values of operating temperatures on the different energetic properties.

\section{Methodology}
\label{sec:methods}
%%%
We propose a multistep process as a method to assess the performance of MOFs for heating and cooling applications and as a tool for finding the optimal working conditions for each adsorbent-fluid working pairs. This method is tested with five MOFs, namely ZIF-8, ZIF-71, ZIF-90, MIL-140C, and MAF-6, for the adsorption of methanol and ethanol working fluids. The multistep process consists of a combination of molecular simulations, thermodynamical modeling, and \textit{in-house} algorithms that iteratively analyze the performance of the AHP/ACS process.

\subsection{Adsorbents}

ZIF-8, ZIF-71, ZIF-90, and MAF-6 are Zn-based Zeolitic Imidazolate Frameworks (ZIFs), and MIL-140C is Zr-based MOF already used for this type of applications.\cite{deLange2015a} Table S1 summarizes the topology and structural properties, pore volume (V$_p$), surface area (SA), framework density ($\rho$), and helium void fraction (HvF) of the selected adsorbents. ZIF-8 \cite{Park2006} and ZIF-90 \cite{Shieh2013} exhibit SOD topology and are formed by Zn metal centers connected to 2-methylimidazole and 2-carboxydehyde organic linkers, respectively. The SOD topology is characterized by sodalite central cages interconnected through small windows. ZIF-8 and ZIF-90 have pores in the range of 10.5-11 Å and apertures of 3.5 Å. ZIF-71 \cite{Liu2013} and MAF-6,\cite{He2015} with RHO topology, are formed by Zn metal centers connected to Zn metal centers connected with 4,5-dichloroimidazole and 2-ethylimidazole organic linkers, respectively. The RHO topology structures are constructed by large central cages interconnected through channels. These two MOFs have similar pore sizes, big cages of about 16.5-17.5 Å and cylindrical channels of 6.5-7 Å. The nature of the organic ligand exerts strong influence in the available pore volume of ZIF-71 and makes MAF-6 the MOF with the largest pore volume. MIL-140C \cite{Guillerm2012} is a five-coordinated Zr-based MOF with a biphenyl-4,4'-dicarboxylic acid linker. The structure has small triangular-shaped channels of about 6 Å that propagates along the c-axis. Of the five structures, MIL-140C is the smallest in terms of available pore volume and surface area. Figure S1 displays the schematic framework connectivity and pore size distribution (PSD). The schematic representation of the ligands can be found in the Electronic Supporting Information, ESI (Figure S2).

\subsection{Simulation Details}

Adsorption capacity as a function of temperature or external pressure was calculated using Monte Carlo (MC) simulations in the Grand-Canonical ensemble (GCMC) with the RASPA software.\cite{Dubbeldam2016, Dubbeldam2013} The MC production run cycles range between 7 $\cdot$ 10$^5$ and 2.2 $\cdot$ 10$^6$. The number of cycles changes depending on the nature of the system working fluid-adsorbent to ensure that the results fluctuate around an equilibrium value.\cite{Luna-Triguero2019} The final data are averaged over the last 5 $\cdot$ 10$^5$ MC cycles.
The interactions between adsorbates and adsorbents are described with Lennard-Jones and Coulombic potentials. We employed Lorentz-Berthelot mixing rules \cite{Boda2008} to calculate Lennard-Jones crossed terms between different atoms, and no tail corrections were applied to the Lennard-Jones. We fixed the length of the simulation box so that it always exceeds twice the spherical cut-off of 12 Å. Electrostatic contributions to the energy of the system have been obtained with the Ewald summation method.\cite{Darden1993} The flexible pseudo-atom model included in the Transferable Potentials for Phase Equilibria (TraPPE) force field \cite{Stubbs2004} was used to describe the alcohol molecules. 

The structure of the adsorbents is considered rigid during the simulations, placing the framework atoms in their crystallographic positions. We used the reported crystal structures of ZIF-8,\cite{Park2006} ZIF-71,\cite{Banerjee2008} MIL-140C,\cite{Guillerm2012} ZIF-90,\cite{Morris2008} and MAF-6.\cite{Huang2006} As it is common in the case of MOFs, the Lennard-Jones parameters for the adsorbents are taken from DREIDING \cite{Mayo1990} except for the metal atoms taken from UFF.\cite{Rappe1992} The partial charges of the adsorbents (see Figure S2 of the ESI) are calculated using the EQeq method\cite{Ongari2019} except for MAF-6, taken from the previous work of Gutierrez-Sevillano \textit{et al.}\cite{Gutierrez-Sevillano2013} Since the interaction between molecules of alcohols via hydrogen bonds plays an essential role in the adsorption mechanism,\cite{Madero-Castro2019} we computed the average number of hydrogen bond per molecule (nHB) using the methodology described in our previous work.\cite{Madero-Castro2019}

\subsection{Thermodynamic Model}

The AHP cycle consists of two steps for adsorption (isobaric adsorption and isosteric heating) and two for desorption (isobaric desorption and isosteric cooling), as represented in Figure 1. The system has different operating conditions pairs (p,T) during the cycle, the temperature of the evaporator (T$_{ev}$), the temperature of the condenser (T$_{con}$), the temperature of desorption (T$_{des}$), and intermediate temperatures (T$_{ev}$ $<$ T$_{con}$ $<$ T$_1$ $<$ T$_2$ $<$ T$_3$ $<$ T$_{des}$), and associated pressures. 
The adsorption and desorption phases of an AHP are characterized by the energy in the different stages: heat taken from the evaporator (Q$_{ev}$), required energy for desorption or regeneration (Q$_{reg}$), heat released by the condensed fluid (Q$_{con}$), and heat released during the adsorption (Q$_{ads}$) at intermediate temperature. For practical reasons, it is usual to assume that T$_1$, also called the minimum temperature of adsorption (T$_{ads}$), is equal to T$_{con}$. 

\begin{figure}
    \centering
    \includegraphics[width=0.46\textwidth]{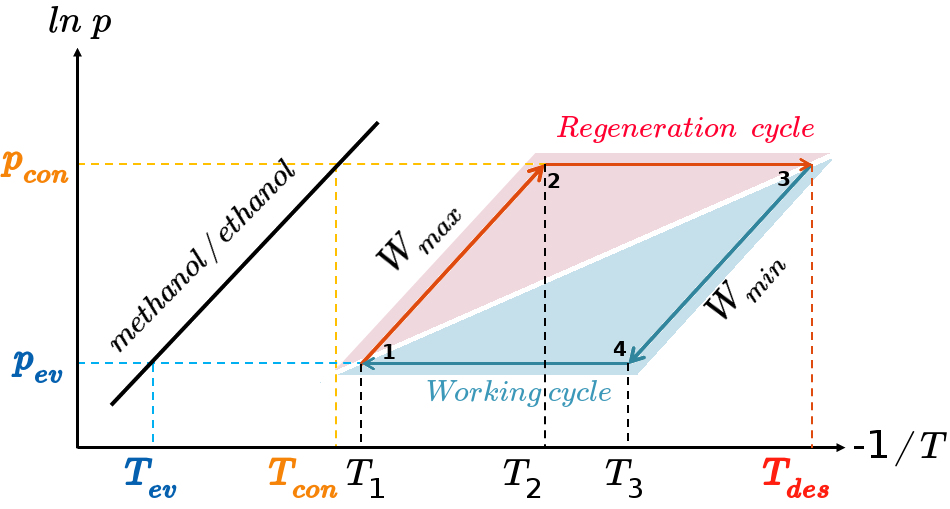}
    \caption{Isosteric cycle of an AHP, including vapor pressure of alcohol molecules (black line), temperature and pressure of the evaporator (T$_{ev}$, p$_{ev}$), and the condenser (T$_{con}$, p$_{con}$), desorption temperature (T$_{des}$), and intermediate cycle temperatures (T$_{(1-3)}$). Highlighted in blue, the working cycle (adsorption) and in red is the regeneration cycle (desorption).}
    \label{fig:fig_01}
\end{figure}

The coefficient of performance (COP), used to describe the energetic efficiency, is defined as the ratio between the obtained energy and the input energy. The obtained energy or delivered energy is also defined as the specific heating or cooling effect (SHE or SCE). For heating, the COP is defined as

\begin{equation}
    COP_H = \frac{-\left( Q_{con} + Q_{ads} \right)}{Q_{reg}} = \frac{SHE}{Q_{reg}}
\end{equation}

\noindent and for cooling

\begin{equation}
    COP_C = \frac{Q_{ev}}{Q_{reg}} = \frac{SCE}{Q_{reg}}
\end{equation}

\noindent Q$_{con}$ and Q$_{ev}$ are proportional to the enthalpy of vaporization ($\Delta$H$_{vap}$), the density of the working fluid $\left( \rho^{wf}\right)$ in confinement, and the working capacity ($\Delta$W), which is the difference between the maximum and minimum adsorption isosteres (W$_{max}$-W$_{min}$). The calculation of the energy needed for regeneration (Q$_{reg}$) and the energy released during adsorption (Q$_{ads}$) requires the estimation of the enthalpy of adsorption, $\Delta$H$_{vap}$. The equations that govern the AHP process from a thermodynamic perspective are detailed in the literature,\cite{deLange2015a}. 

The enthalpy of vaporization is taken from Majer \textit{et al.},\cite{Majer1986} while the remaining parameters are calculated using models. Based on the Dubinin-Polanyi (DP) theory,\cite{Lehmann2017,Lehmann2017a}  any equilibrium adsorption curve where loading is pressure and temperature dependent, q(p,T), can be reduced to a characteristic curve. The characteristic curve is the relation between the potential of adsorption (A) and the adsorbed specific volume (W) defined as:

\begin{equation}
A=RT \left( \ln \frac{p_0 \left(T\right)}{p} \right)
\end{equation}

\begin{equation}
W=\frac{q\left(p,T \right)}{\rho^{wf} \left( T \right)}
\end{equation}

\noindent where p$_0$ is the vapor pressure of the working fluid, q is the mass adsorbed, and $\rho^{wf}$ is the density of the working fluid in the adsorbed phase. We used the Peng-Robinson equation of state \cite{Peng1976} to set the saturation pressure. The DP theory allows to calculate adsorption enthalpy from the characteristic curve using numerical techniques. We calculated the adsorbate density using the Hauer model.\cite{Nagel2015} This model establishes linear relations between the bulk density and the density inside the pores of the structure. The model was initially developed for water and has been modified to estimate the density of alcohols where $\rho$(T$_0$) is the free liquid density taken from experimental data at the reference temperature (T$_0$=283 K).\cite{Gonzalez2007}

\begin{figure*}[t!]
    \centering
    \includegraphics[width=\textwidth]{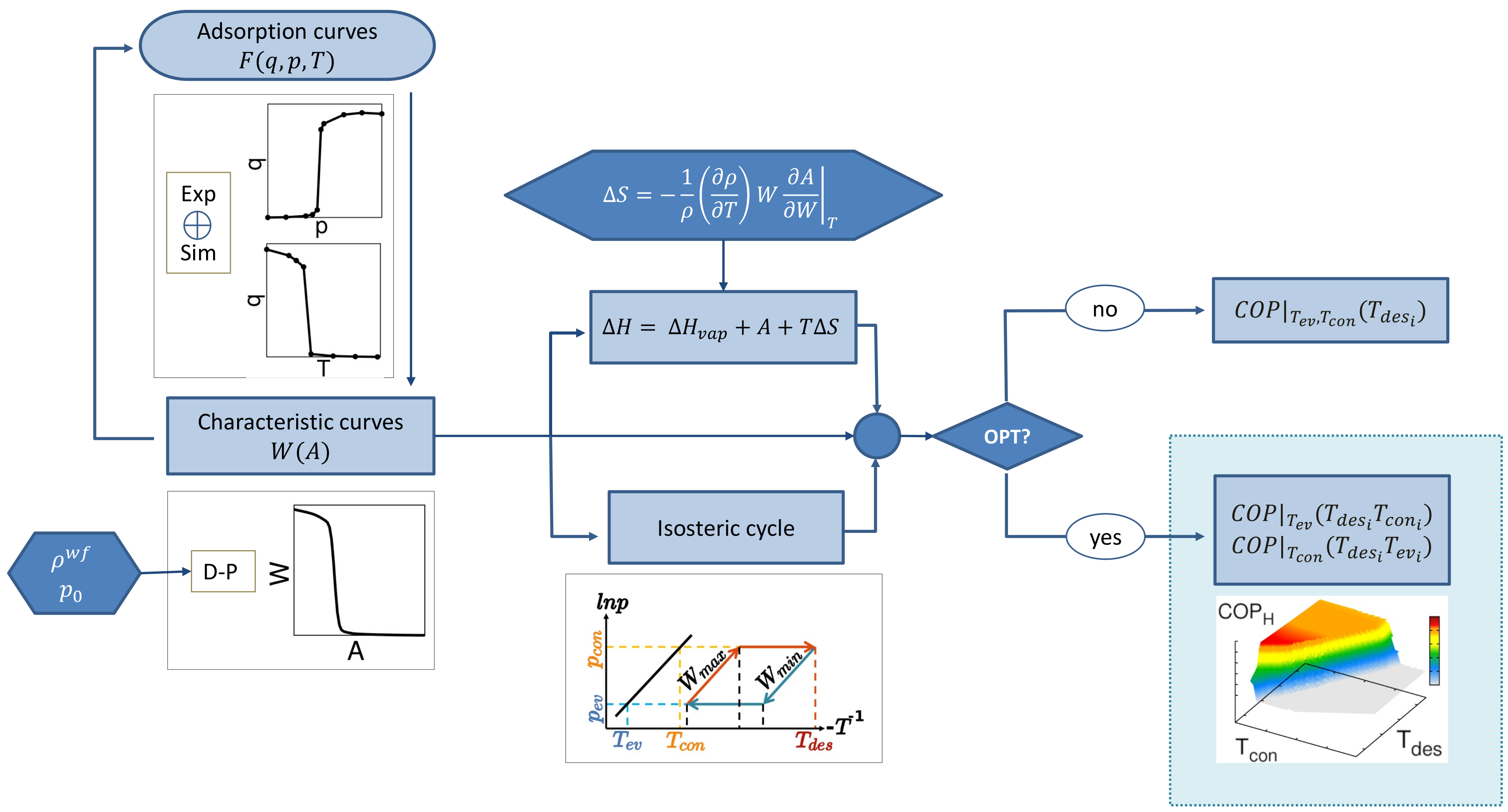}
    \caption{Schematic algorithm for the multistep process.}
    \label{fig:fig_02}
\end{figure*}

\begin{equation}
\rho^{wf} \left(T\right)=\rho\left( T_0 \right) \left[ 1-\alpha_T \left( T-T_0 \right) \right]
\end{equation}

The coefficient of thermal expansion, $\alpha_T$, is considered constant for each working fluid. $\alpha_T$ has been calculated using the equation

\begin{equation}
\alpha_T=-\frac{1}{\rho\left( T \right)} \left( \frac{\partial \rho \left( T \right)}{\partial T} \right)_p
\end{equation}

\noindent using the density of the fluid at high pressure, 100 MPa. At T = 300 K, $\alpha_T$ is 8.026 $\cdot$ 10$^{-4}$ [1/K] for methanol and 7.285 $\cdot$ 10$^{-4}$ [1/K] for ethanol.

\subsection{Multistep Process}

The process is schematically represented in Figure 2. \textbf{Step 1} is to obtain the equilibrium adsorption curves, e.g., adsorption isotherms and isobars. This can be done either experimentally or with GCMC simulations. In \textbf{Step 2} the input data are processed according to Dubinin-Polanyi theory. This leads to a temperature invariant characteristic curve that allows i) to predict new adsorption curves at any conditions (step 1)  and ii) to proceed with the thermodynamic analysis of the system. 

The application of the DP theory requires inserting pre-processed functions into the algorithm. The most important variables are the density of the fluid in confinement $(\rho^{wf})$ and the saturation pressure (p$_0$). Once the characteristic curve has been obtained, the process bifurcates. During \textbf{Step 3}, we calculate the enthalpy of adsorption of the system, which depends on the enthalpy of evaporation of the fluid, adsorption potential, and entropy. The entropy calculation is pre-processed due to its dependency on the thermal expansion coefficient of the fluid in confinement. In \textbf{Step 4}, the isosteric cycle is calculated, estimating the maximum and minimum capacity isosteres for given conditions, i.e., setting T$_{ev}$ and T$_{con}$ and the associate pressures. The calculation of COP for either heating or cooling can be done under different conditions (\textbf{Step 5}). Without optimization, the temperature of the evaporator and the condenser are fixed, and therefore the associated pressures. The COP is assessed as a function of the desorption temperature. \textbf{Step 6} is the optimization step. Here we scan the operating conditions and associated thermodynamic parameters in subsequence loops. The variation of T$_{ev}$, T$_{con}$, and T$_{des}$ are carried out simultaneously, with the only condition of T$_{ev}$ < T$_{con}$ < T$_{des}$. After the massive data generation, the algorithm computes the COP by setting only one temperature (T$_{ev}$ or T$_{con}$) and scanning for the optimal conditions. From this step, a 3D plot per fixed variable is obtained, which allows to choose the operating conditions that provide the maximum performance. Using the same procedure, we calculated other quantities of interest, such as working capacity, specific heating effect (SHE), specific cooling effect (SCE), and heat released to the condenser (Q$_{con}$) or the evaporator (Q$_{ev}$). 

\section{Results and Discussion}
\label{sec:results}

\begin{figure*}[t!]
    \centering
    \includegraphics[width=0.8\textwidth]{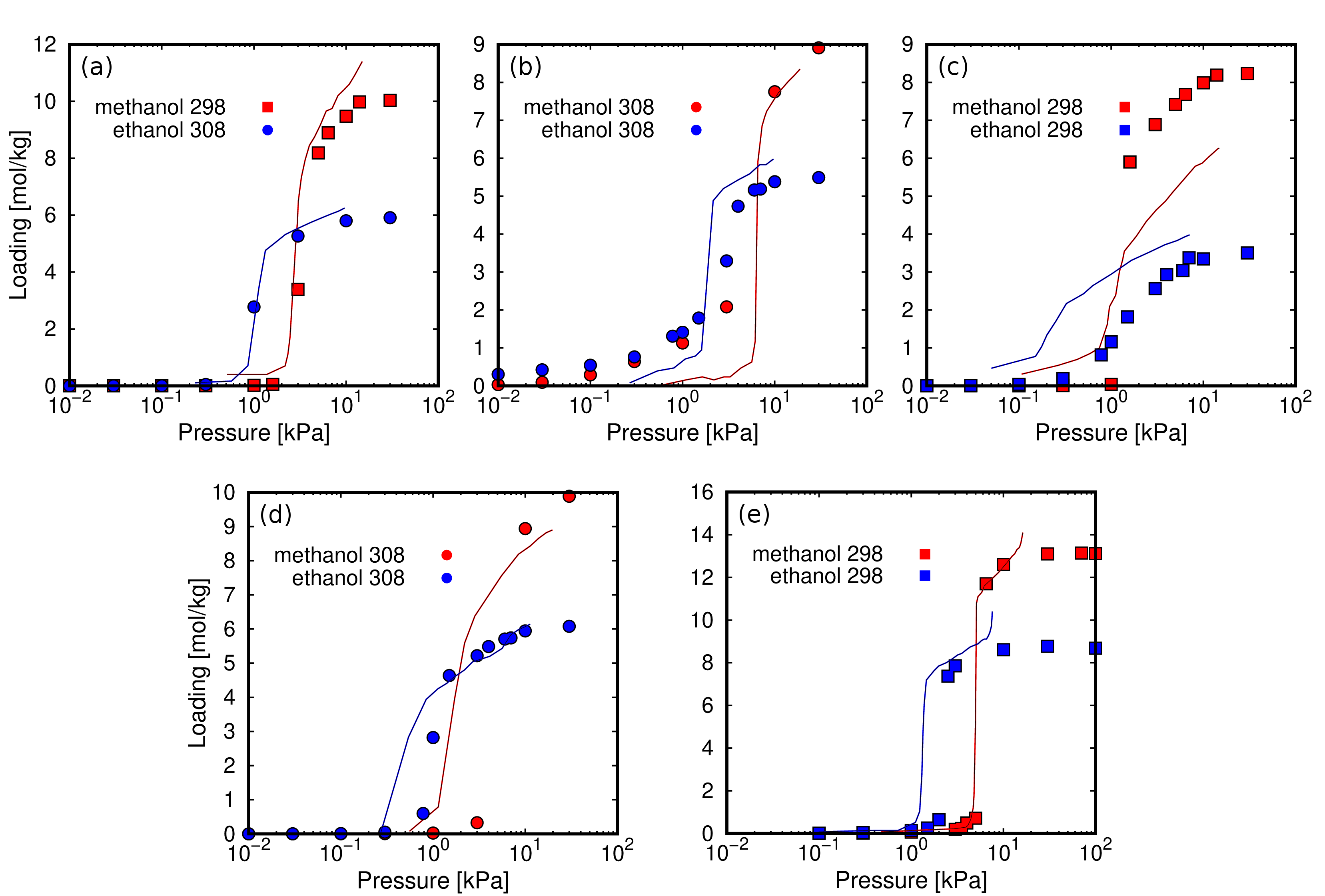}
    \caption{Computed adsorption isotherms (symbols) of methanol (red) and ethanol (blue) in (a) ZIF-8, (b) ZIF-71, (c) MIL-140C, (d) ZIF-90, and (e) MAF-6 at 298 K (squares) and 308 K (circles). The experimental values (solid lines) are taken from the literature (ZIF-8,\cite{deLange2015a,Zhang2013} ZIF-71,\cite{Zhang2013} MIL-140C,\cite{deLange2015a} ZIF-90,\cite{Zhang2013} and  MAF-6.\cite{He2015a}).}
    \label{fig:fig_03}
\end{figure*}

We calculated the adsorption isotherms of methanol and ethanol to compare with reported experimental data for ZIF-8,\cite{deLange2015a,Zhang2013} ZIF-71,\cite{Zhang2013} MIL-140C,\cite{deLange2015a} ZIF-90,\cite{Zhang2013} and  MAF-6.\cite{He2015a} This first set of calculations (Figure 3) was used to test the suitability of the force field to reproduce the experimental adsorption of these polar molecules in the selected MOFs. Computed and experimental adsorption isotherms are in line. Discrepancies in the onset pressures of ZIF-90 and ZIF-71 are attributed to the use of a generic force field, which has not been refitted for a particular adsorbent-adsorbate pair. For example, ZIF-90 has exposed oxygen atoms in the ligand, while ZIF-71 has chlorine atoms accessible to the adsorbates. It is expected that alcohol molecules interact through the -OH groups with these electronegative atoms. However, the generic force field is not trained to reproduce these particular interactions and can cause a slight overestimation and underestimation of the onset pressures of ZIF-71 and ZIF-90, respectively. We found larger deviations for the adsorption isotherms of the two alcohols in MIL-140C, overestimating the adsorption capacity of methanol and underestimating the adsorption capacity of ethanol. Based on that, we cannot rule out any flexibility effect of the MOF, neglected in the simulations. However, the general behavior of all isotherms is well described by the selected force field, making it suitable for studying the adsorption of alcohols in zeolitic imidazolate frameworks.

Figure 4 compares the adsorption characteristics of the working fluids at room temperature. The adsorption isotherms of methanol and ethanol in ZIF-8 and ZIF-90 (Figures 4a and 4b), which have SOD topology, show very similar behavior. This can be explained in terms of similar structural properties and composition. The other pair of adsorbents that share RHO topology are ZIF-71 and MAF-6. The presence of chlorine atoms in ZIF-71 has significant effects on the adsorption. Compared to MAF-6, we observe a decrease in the adsorption capacity and a slight attenuation of the abrupt step caused by the higher hydrophilicity.  As the most hydrophilic MOF, MAF-6 shows the highest adsorption capacity for both adsorbates. MIL-140C shows the lowest ethanol capacity, while the adsorption of methanol is like that obtained in the other MOFs. This is due to a more efficient packing of methanol than ethanol within the small channels of MIL-140C, which is directly related to kinetic diameters, 3.6 and 4.3 Å, respectively.\cite{Tang2019}

\begin{figure}[t!]
    \centering
    \includegraphics[width=0.48\textwidth]{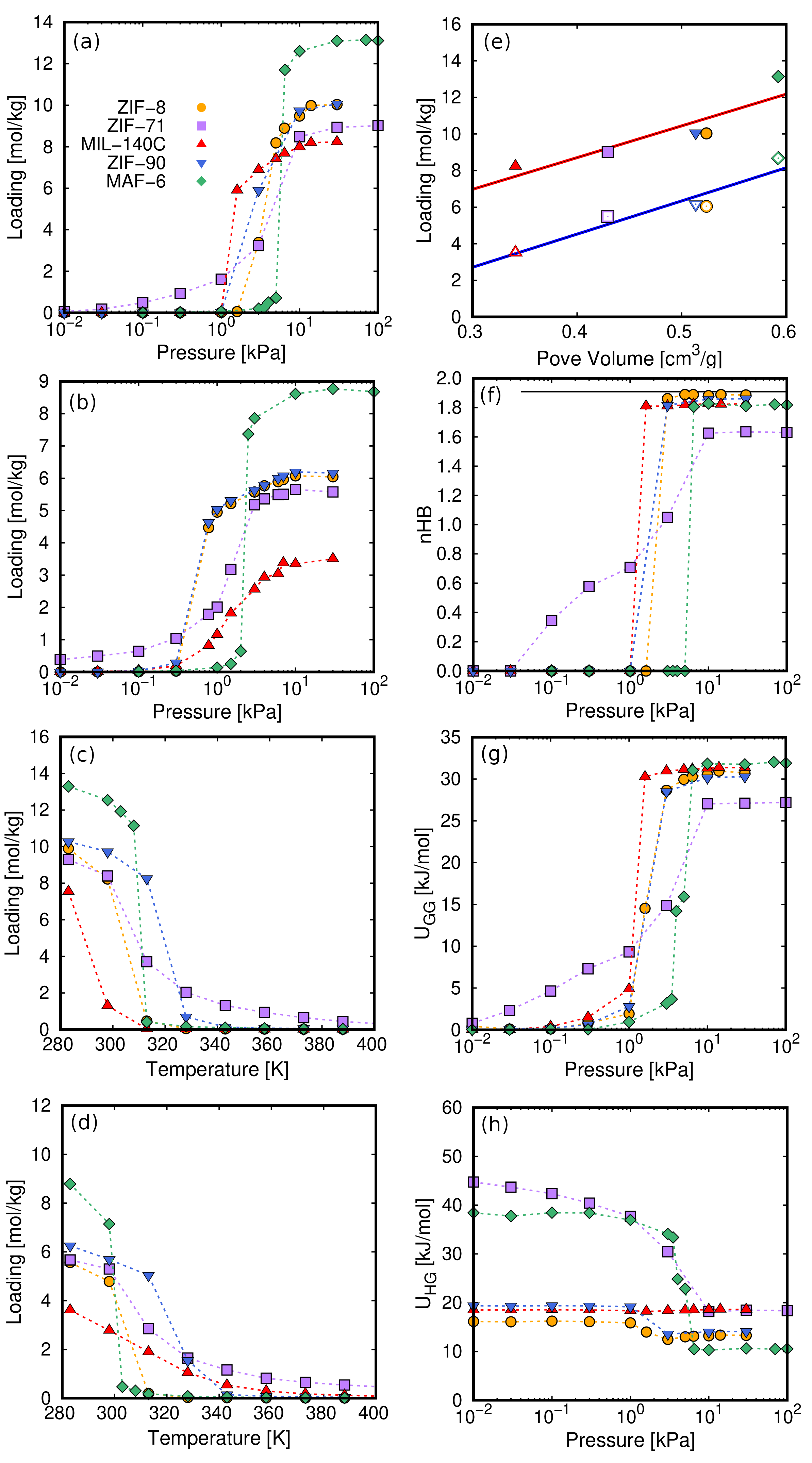}
    \caption{Adsorption properties of methanol and ethanol in the selected MOFs. Computed adsorption isotherms of (a) methanol and (b) ethanol at 298 K, adsorption isobars of (c) methanol, and (d) ethanol at the selected working pressures shown in Table S2 (ESI). (e) Saturation capacity of methanol (closed symbols) and ethanol (open symbols) as a function of pore volume. The solid lines in (e) stand for the fitted values to a straight line showing the dependence of loading with the pore volume. (f) nHB, (g) guest-guest interactions, and (h) host-guest interactions of methanol as a function of the pressure. Non-depicted error bars denote fluctuations smaller than the symbol size.}
    \label{fig:fig_04}
\end{figure}

For practical application, it is important to maximize the thermodynamic efficiency of the system.\cite{Aristov2013} The choice of operating conditions is key in the design of AHP and ACS devices. As shown in Figure 1, the thermodynamic cycle involves two isobaric and two isosteric steps. Hence, it is convenient to analyze the adsorption isobars in addition to the adsorption isotherms described. From an energetic point of view, we are interested in an adsorption isobar with a single steep step. It is known that small variations in the regeneration conditions can lead to large changes in performance.\cite{Li2019} In this regard, we use pressure control as a mechanism to improve efficiency. The working pressure is based on the onset pressure of the adsorption isotherm at room temperature. The pressure is set for each working pair as the lowest value of pressure that ensures high uptake (immediately after the step in the adsorption isotherm at 298 K). All temperature and pressure conditions of the adsorption isotherms and isobars calculated in this work can be found in the ESI (Table S2). 

Figures 4c and 4d show the adsorption isobars of methanol and ethanol in the selected MOFs. These isobars make it possible to determine the regeneration temperature for each working pair under the selected operating conditions. The slope in the desorption isobar is a first indication of the efficiency and performance of a particular working pair. The temperature window for the desorption process is in the range of 300 - 340 K. Above this range, all structures have released most of the methanol and ethanol load. The maximum adsorption capacity can be related to the pore volume of the adsorbent (Figure 4e) with the following trend: MAF-6 $>$ ZIF-8 $\simeq$ ZIF-90 $>$ ZIF-71 $>$ MIL-140C.

The adsorption isotherms calculated for ZIF-8, ZIF-90, and MAF-6 have a steeped behavior, less prominent for ZIF-71 and MIL-140C. This is due to the strong guest-guest interactions of the adsorbates inside the pores, a typical characteristic of hydrophobic materials. These interactions are driven by hydrogen bonds (HB). Figure 4f shows the average number of hydrogen bonds per molecule (nHB) as a function of pressure for methanol. The values for ethanol can be found in the ESI (Figure S3a). In the onset pressure where the adsorption occurs, we found an abrupt increase in the nHB values. At high loading, the nHB in confinement is similar to that in the bulk.\cite{Madero-Castro2019} We found exceptions for methanol and ethanol in ZIF-71 and ethanol and in MIL-140C. The type of organic ligand is the cause of this situation in ZIF-71. The chlorine atoms are partially located in the large cage, which decreases the degree of hydrophobicity of the structure. This is reflected in the shape of the adsorption isotherm and a lower number of hydrogen bonds within the cavities. MIL-140C is formed by channels instead of by cages. This particular topology, as well as the size of the structure, motivates that the number of hydrogen bonds per ethanol molecule is lower than in the rest of the MOFs.  

The structure of the hydrogen bonds changes in confinement, with a competition between molecules of alcohol with one or two HBs (see Figure S4 in ESI), and this is reflected in the guest-guest potential energy (U$_{GG}$) as a function of external pressure (see Figure 4g for methanol and ESI, S3b, for ethanol). All these findings evidence that the adsorption mechanism is driven by the nucleation of the polar molecules through hydrogen bond interactions. This behavior is similar to that found in the adsorption of other polar compounds, such as ammonia, on adsorbents with large cavities.\cite{Matito-Martos2020} 

\begin{figure*}[t!]
    \centering
    \includegraphics[width=0.8\textwidth]{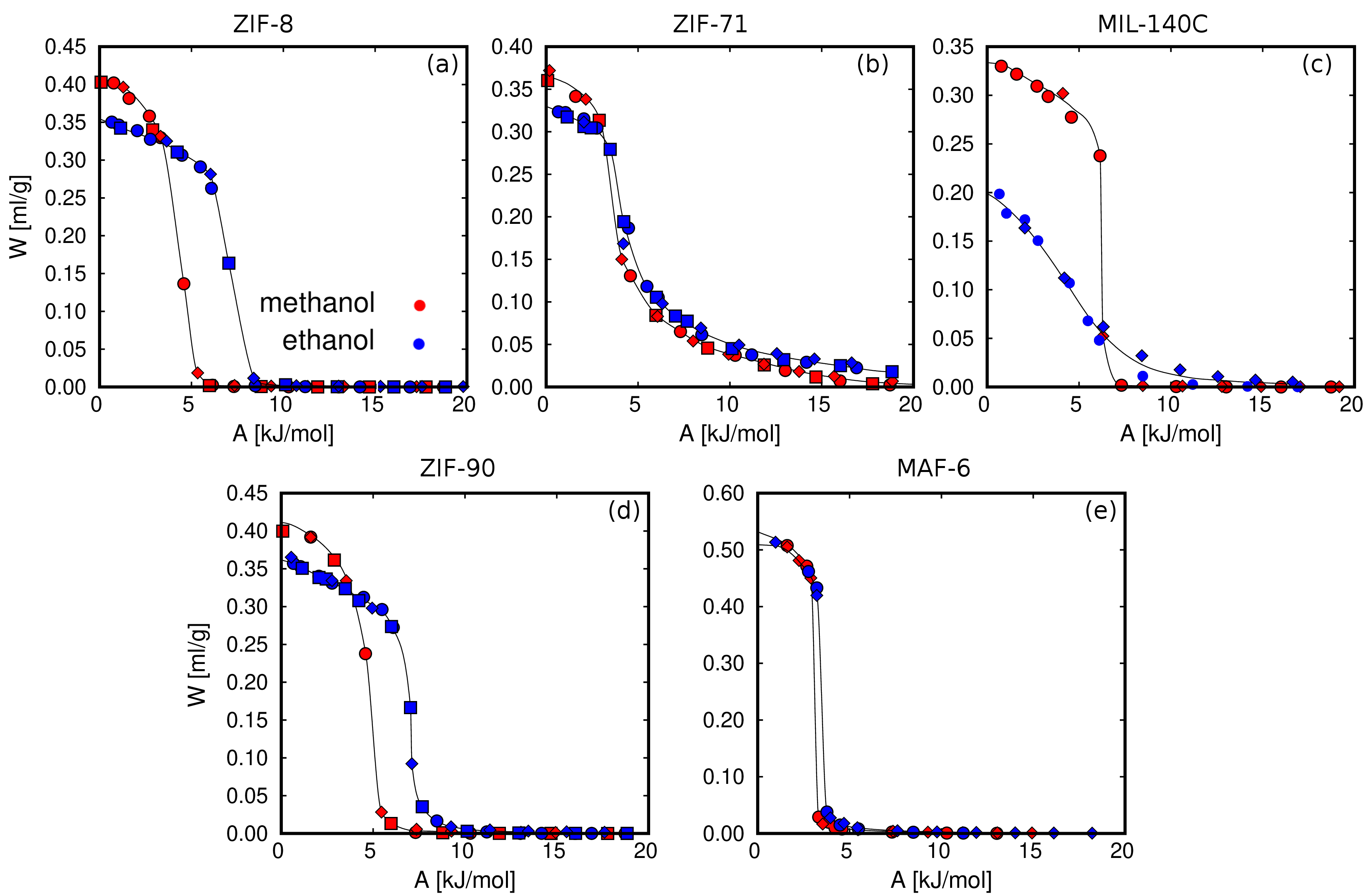}
    \caption{Characteristic curves determined from the calculated adsorption isotherms and isobars (a) ZIF-8, (b) ZIF-71, (c) MIL-140C, (d) ZIF-90, and (e) MAF-6. The lines correspond to fit curves obtained by splines.}
    \label{fig:fig_05}
\end{figure*}

The energy contribution corresponding to the interaction between methanol and adsorbents is shown in Figures 4h (for ethanol, see ESI, Figure S3c). The affinity between the adsorbates and the internal surface of the framework increases with the host-guest energy. Host-guest interactions generally weaken with loading as preferential adsorption sites fill up and guest-guest interactions become more important. The adsorption onset pressure is strongly related to the host-guest interactions,\cite{Li2009} surface area of the structure, and the kinetic diameter of the adsorbate. In Figure 4, the infinite dilution regime corresponds to values of pressure below the step in the adsorption isotherm. This regime leads to a reduction in host-guest energy for all the structures except MIL-140C. The channel-like structure of this relatively small MOF hinders the nucleation of alcohol molecules and leads to nearly constant values in the host-guest energy for the two adsorbates over the entire pressure range.

\begin{figure}[t!]
    \centering
    \includegraphics[width=0.305\textwidth]{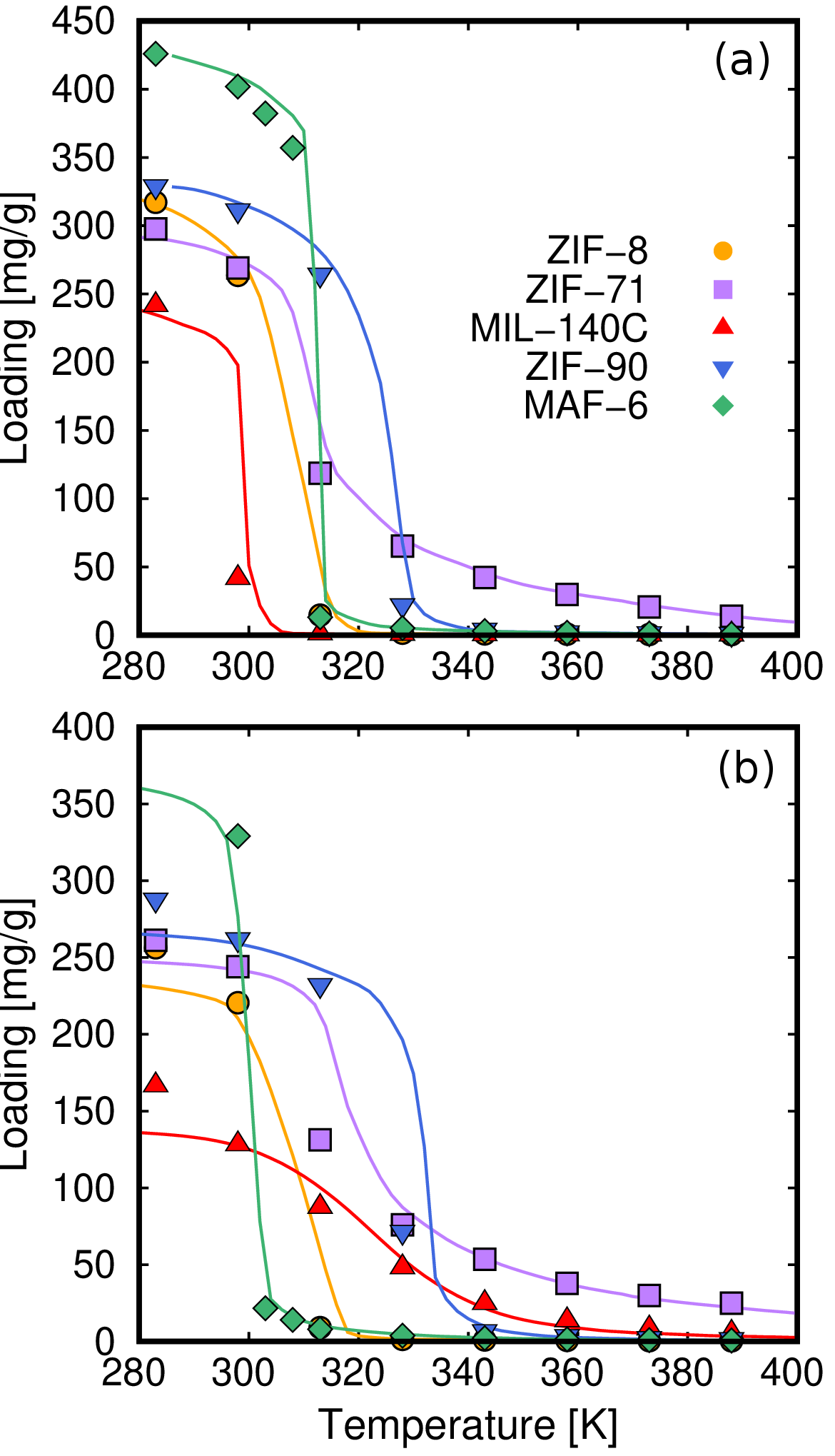}
    \caption{Adsorption isobars calculated with GCMC simulations (symbols) and from the characteristic curve (lines) of methanol (a) and ethanol (b) in all the adsorbents at the selected working pressures (Table S2).}
    \label{fig:fig_06}
\end{figure}

\begin{figure*}[t!]
    \centering
    \includegraphics[width=0.8\textwidth]{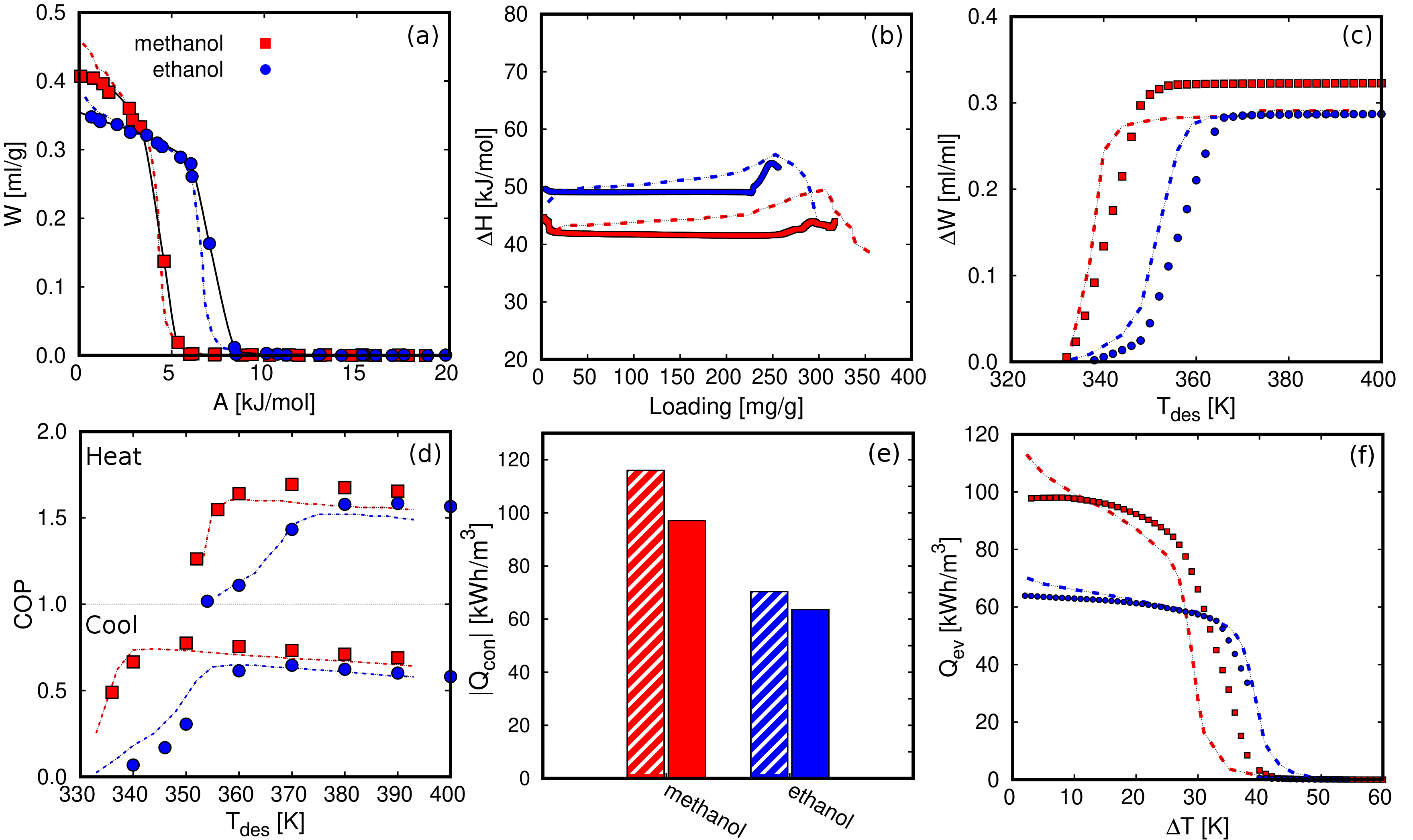}
    \caption{Comparison between experimental (dashed lines) \cite{deLange2015a} and computational results (solid symbols) of methanol and ethanol in ZIF-8 for (a) characteristic curves, (b) heat of adsorption as a function of the loading, (c) deliverable capacity ($\Delta$W) as a function of the desorption temperature for refrigeration conditions, T$_{ev}$ = 278 K and T$_{con}$ = 303 K, (d) COP as a function of the desorption temperature for heating and cooling, (e) Q$_{con}$ at T$_{con}$ = 298 K, and (f) Q$_{ev}$ as a function of the temperature lift (T$_{con}$ - T$_{ev}$), varying T$_{ev}$ with T$_{con}$ and T$_{des}$ set to 298 K and 373 K, respectively.}
    \label{fig:fig_07}
\end{figure*}

The low coverage host-guest energy is approximately twice as high for ZIF-71 and MAF-6 as for ZIF-8 and ZIF-90, due to the interaction of the first molecules entering the structures with specific binding sites. Methanol and ethanol first interact with the exposed chlorine atoms of the dichloroimidazole link of ZIF-71. As for MAF-6, the alcohol molecules interact with the aromatic rings of the ligand by electrostatic interactions, where the hydroxyl group aligns with the electrostatic field lines pointing towards the center of the ring.\cite{Madero-Castro2019} After the step in the adsorption isotherm, the host-guest energy decreases to a constant minimum value, while the guest-guest energy reaches the maximum value in the same pressure range. This transition from host-guest to guest-guest interactions points to the nucleation of molecules in hydrophobic MOFs with large pores. Similarly, this energy exchange is the origin of the energy released during the adsorption process, which governs the AHP and ACS devices.

The next stage of our multistep process is the post-processing of the adsorption data using the thermodynamic model. Here, the characteristic curve of adsorption is critical as we can extract a set of relevant quantities from it. As described in the methods, the DP theory relates the adsorption potential (A) and the amount of adsorbed volume (W). One of the limitations is the assumption of the temperature invariance of W. To ensure the applicability of the DP theory, all calculated adsorption isotherms and isobars (see Table S2 of the ESI) must converge to the same characteristic curve. Figure 5 shows the characteristic curves of the working pairs from all calculated adsorption curves. The agreement of the different transformations indicates the suitability of the DP theory in the systems under study. We can then use the characteristic curve to calculate any adsorption equilibrium relation for a given operating condition since any combination of (p,T) is related via the adsorption potential to the working volume and, therefore, to the loading. This interesting property of the characteristic curve can be used to predict adsorption isotherms at different temperatures and adsorption isobars for each working pair. Figure 6 shows the comparison between the adsorption isobars computed with the GCMC simulations and those predicted from the characteristic curve. The prediction of adsorption isotherms at different temperatures can be found in the Figure S5 of the ESI. The results confirm the predictive power of a single characteristic curve, which can be used to relate the adsorption amount with the operating conditions of the thermodynamic cycle.

To explore the limits of our approach, we computed the relevant properties of AHP and ACS devices. Figure 7 compares our results for ZIF-8 with previous data from de Lange \textit{et al.} \cite{deLange2015a} For both working fluids, the characteristic curves are similar in shape and maximum capacity (Figure 7a). The small differences are due to the choice of density, taken as an approximation by de Lange \textit{et al.} \cite{deLange2015a} and using Hauer's model. Figure 7b shows the enthalpy or heat of adsorption as a function of loading. The experimental $\Delta$H was obtained using the Clausius Clapeyron method \cite{Pan1998,Cimino2017} while we are using the Dubinin-Polanyi theory. Figure 7c exhibits the effect of the desorption temperature on the deliverable capacity for refrigeration conditions. As can be seen, our computed results agree with the experiments. Figure 7d depicts the COPs for the two working fluids using the operating conditions specified by de Lange \textit{et al.} \cite{deLange2015a} That is, T$_{ev}$ = 278 K and T$_{con}$ = 303 K for ACS and T$_{ev}$ = 288 K and T$_{con}$ = 318 K for AHP. The agreement is very good, with only a small deviation due to the $\Delta$H calculation method, saturation pressure estimation, or the adsorbed density model. It is worth noting that de Lange \textit{et al.} \cite{deLange2015a} approximates the adsorbed density as the liquid density of the fluid, while we used a linear model to describe the temperature dependence of the density of the fluids in confinement. Figure 7e shows the heat released in the condenser (Q$_{con}$) with T$_{con}$ = 298 K and assuming maximum desorption. Although we used different methods, we found a reasonable agreement. Similarly, Figure 7f shows the heat released in the evaporator as its temperature differs from the condenser temperature. Based on these results, we can assume that the methodology used could be easily extrapolated to the study of porous materials for AHP and ACS applications.

\begin{figure}[t!]
    \centering
    \includegraphics[width=0.46\textwidth]{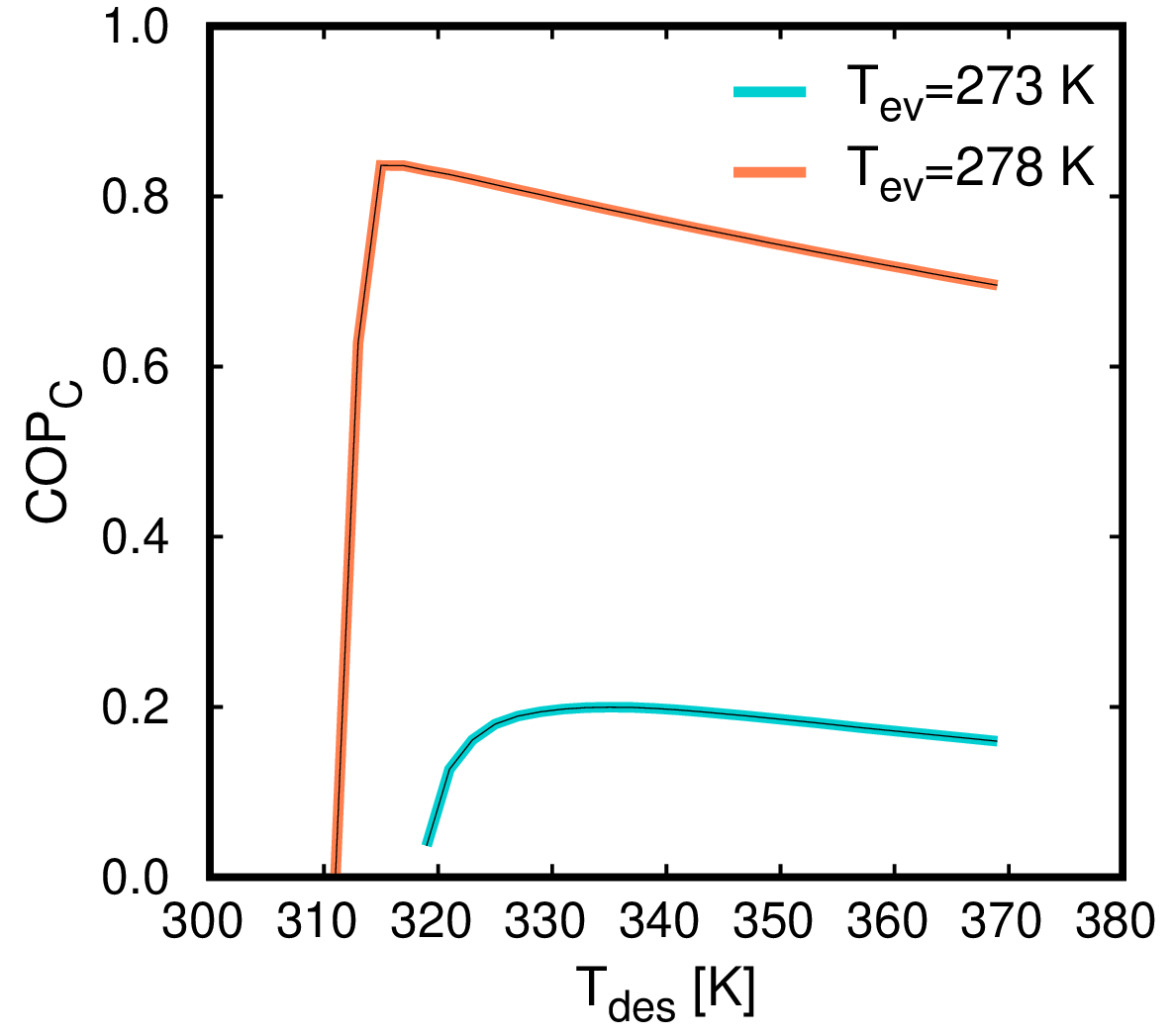}
    \caption{Coefficient of performance (cooling) of ethanol in MAF-6 as a function of desorption temperature, setting the temperature of the evaporator at 273 K and 278 K, respectively. The temperature of the condenser is set to T$_{con}$ = 288 K.}
    \label{fig:fig_08}
\end{figure}

\begin{figure*}[t!]
    \centering
    \includegraphics[width=0.7\textwidth]{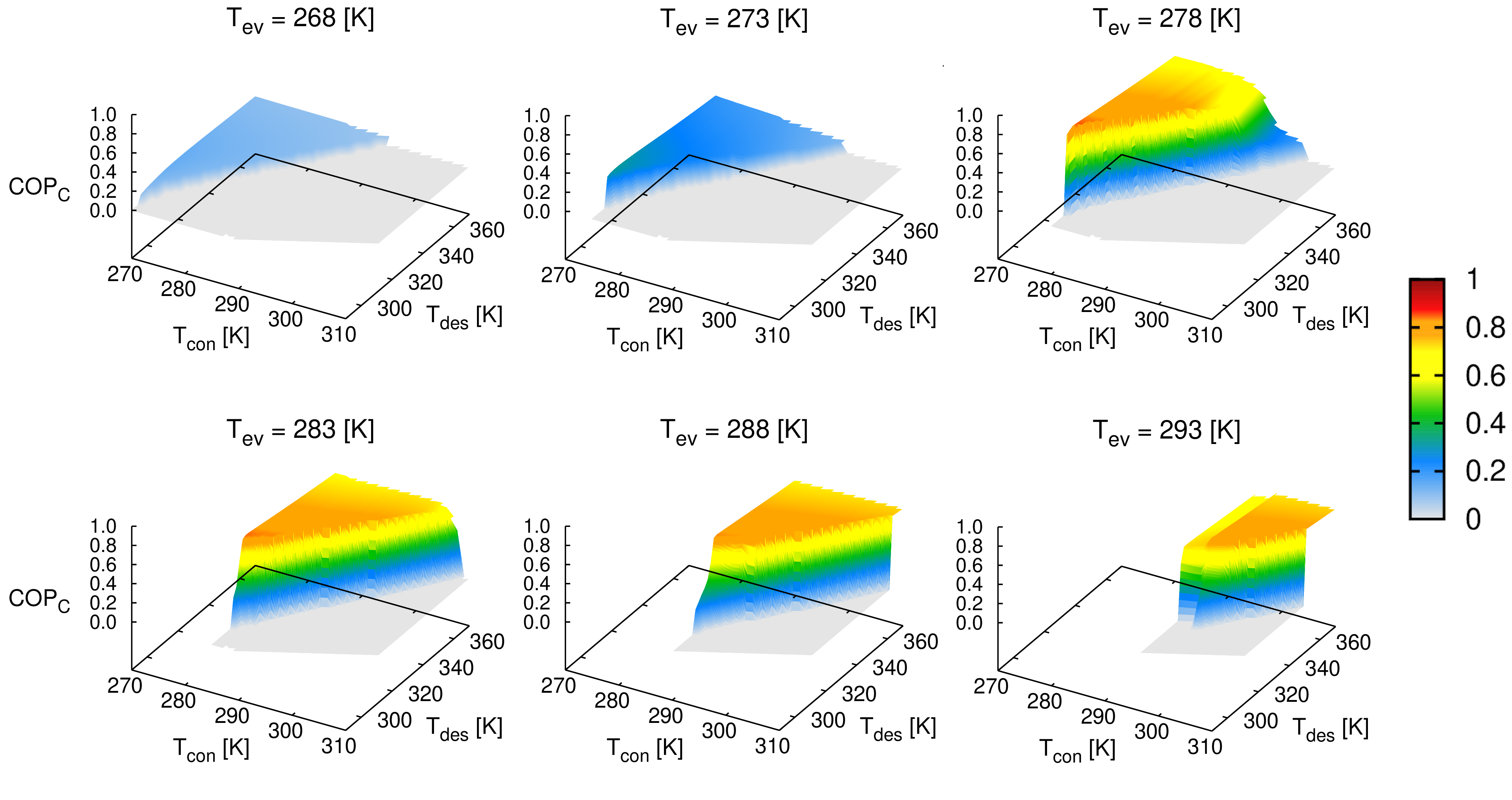}
    \caption{Evolution of the coefficient of performance (cooling) of ethanol in MAF-6 by varying the operating temperature of the thermodynamic cycle (see Figure 1).}
    \label{fig:fig_09}
\end{figure*}

\begin{figure}[t!]
    \centering
    \includegraphics[width=0.4\textwidth]{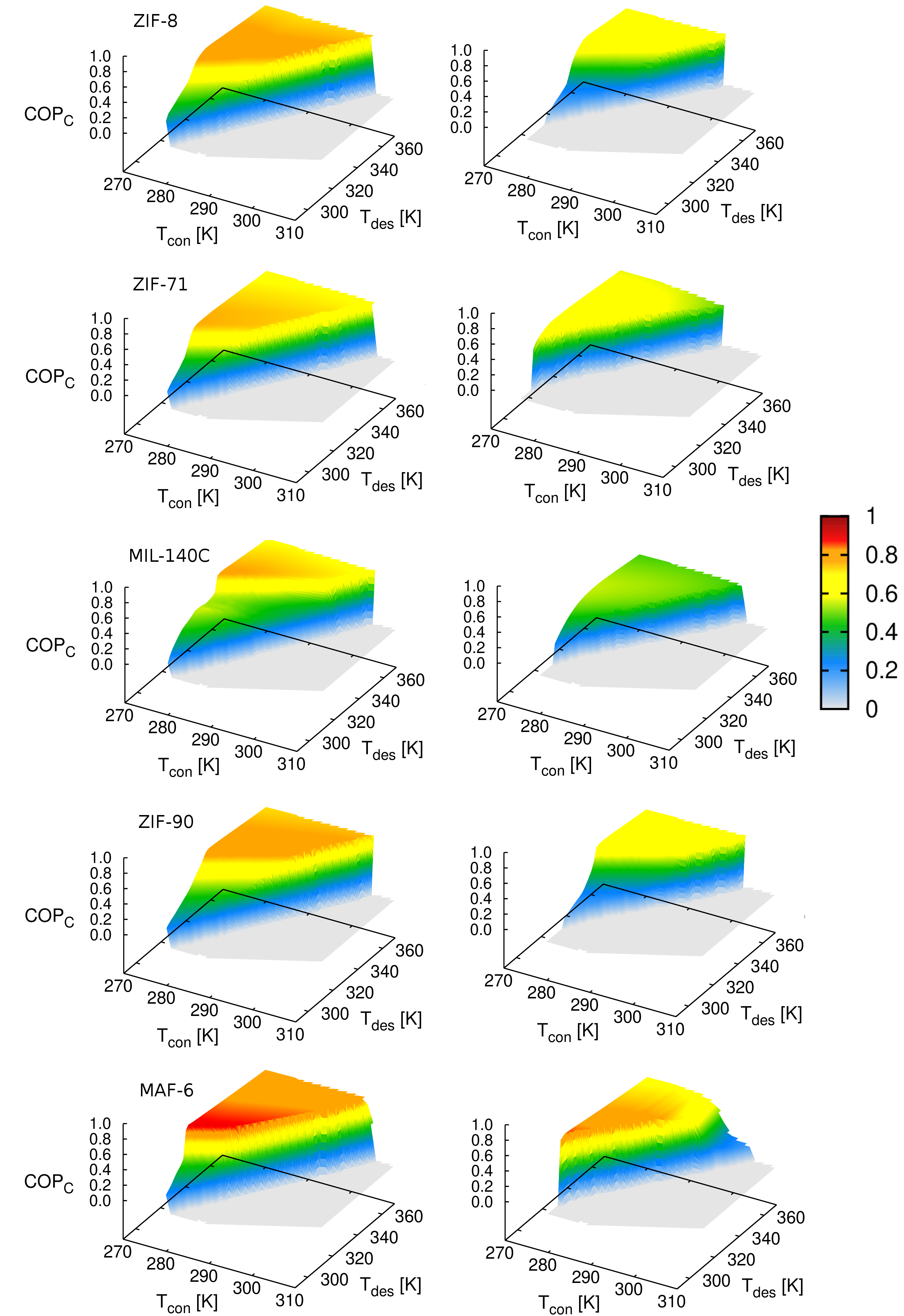}
    \caption{Coefficient of performance (cooling) of methanol (left column) and ethanol (right column) by varying the operating temperature of the thermodynamic cycle (see Figure 1). The temperature of the evaporator is set to T$_{ev}$ = 278 K.}
    \label{fig:fig_10}
\end{figure}

The COP for the cooling and heating processes using methanol and ethanol allows comparing the performance of the adsorbents. In Figure 7d, we can observe that the COP increases to a maximum value at a certain temperature depending if the conditions are for heating or cooling applications. After this point, the COP decreases linearly as temperature increases. Methanol generally shows a slightly higher value than ethanol over the entire temperature range. This is a typical representation of the COP values, that is, varying the desorption temperature but setting the temperatures of the evaporator and condenser, respectively. As this property strongly depends on the selection of the operating conditions (See Figure 1), we can find temperature ranges resulting in low values of COP. However, a single value of the COP could not be an adequate indication of the performance of the material but rather a consequence of the setting of the operating conditions. To clarify this issue, Figure 8 shows the COP for cooling of ethanol in MAF-6, setting the temperature of the condenser to 288 K, and for two nearby values of the temperature of the evaporator, 273 and 278 K, respectively. We can see that a variation of only five degrees in the operating conditions drastically changes the performance of the working pair moving from a maximum COP of 0.2 to 0.85.

The effect shown in Figure 8 is more noticeable for materials depicting steeped adsorption isotherms or isobars. However, steep adsorption is desired for these applications when complete regeneration of the thermodynamic cycle occurs in a small increase of temperature or pressure lift. To avoid misinterpretation of the performance analysis, we suggest deeply examining each working pair and reducing the number of fixed parameters. To do so, we used the proposed multistep process described in the Methodology (see Figure 2), where we iteratively calculated the targeted properties for all possible working temperatures. Thus, instead of having a single relationship between the COP and desorption temperature (Figure 8), we obtain a complex data set that shows the evolution of the COP. This is performed by simultaneously varying all the operating temperatures of the thermodynamic cycle.

Figure 9 represents the COP for cooling of ethanol in MAF-6 as a function of the temperature of the condenser, evaporator, and desorption. From this data set, we can extract the values of the range of operating conditions that maximize the COP. The optimal desorption temperature must be chosen to ensure complete regeneration of the material, avoiding unnecessary overheating and wasted energy. In this case, the ethanol adsorption in MAF-6 shows a maximum COP of 0.84-0.86 at T$_{ev}$ = 275-285 K, T$_{con}$ = 278-288 K, and T$_{des}$ = 300-315 K for ACS applications.

To compare the results obtained in the five MOFs, Figure 10 shows the COP for cooling, setting the temperature of the evaporator to 278 K. This temperature is within the range of operating conditions for cooling applications, and all working pairs reach the optimum level of COP values. In general, methanol performs better than ethanol for cooling applications, which could be due to better molecular packaging leading to higher adsorption capacity. The superior performance of MAF-6 compared to the other adsorbents is remarkable. The high adsorption capacity and especially steeped adsorption behavior make this MOF outperform the rest under study for AHP and ACS applications.

The heat energy transferred to the condenser (Q$_{con}$) that we show in Figure 7e is another relevant property. The values obtained for the five MOFs for both methanol and ethanol are shown in Figures 11a and b. The volumetric energy released in the condenser shows similar values for heating and cooling applications and shows that methanol releases more energy than ethanol. Again, the exception to the rule is the value of ethanol in MAF-6 for cooling applications for the same reasons as for COP. We obtained the Q$_{con}$ assuming a complete regeneration of the cycle, i.e., maximum desorption of the adsorbates. This means that the only temperatures that affect the calculation of Q$_{con}$ are those of the evaporator and condenser. To get a broader overview of the variation of Q$_{con}$ of methanol and ethanol with operating temperatures, we plot Q$_{con}$ as a function of T$_{con}$ and T$_{ev}$ in Figures 11c and d. In addition, Figures S6-S9 show a comparison of the Q$_{con}$ for both working fluids in all the adsorbents. The figures are represented in volumetric (kWh/m$^3$) and gravimetric (kJ/kg) units to show the effect of the framework density on the energy released by the adsorbent. From all the data, we selected MAF-6 as the reference MOF because it was the best performing structure. In line with the adsorption isotherms and isobars, MAF-6 shows a pronounced Q$_{con}$, suggesting the importance of selecting the working conditions. After the step, all curves converge to a similar maximum Q$_{con}$ value, which is the optimal value for each working pair.

\begin{figure}[t!]
    \centering
    \includegraphics[width=0.48\textwidth]{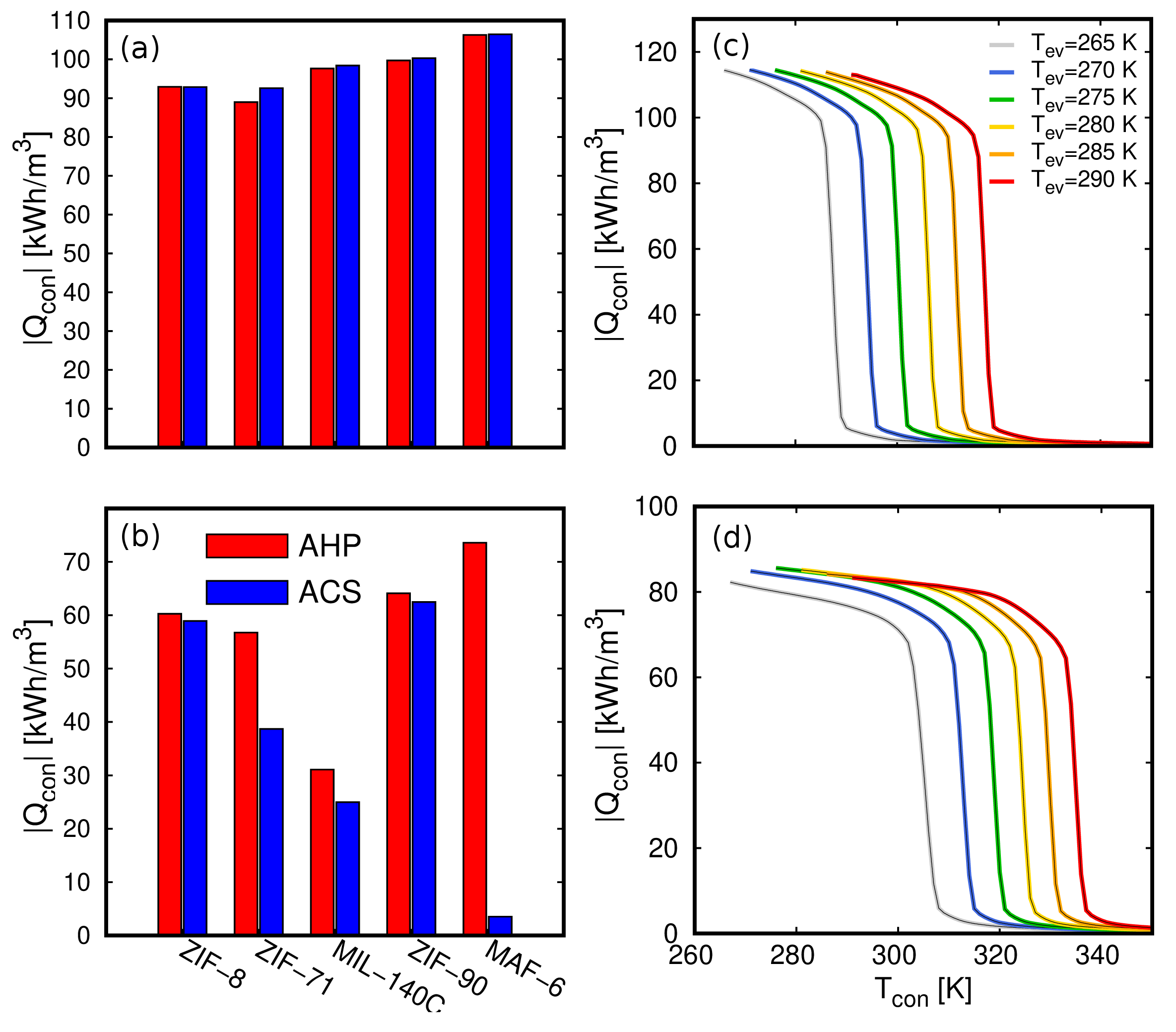}
    \caption{Volumetric heat energy transferred to the condenser per unit of volume of MOF using (a) methanol and (b) ethanol for AHP (red) with T$_{ev}$ = 273 K and T$_{con}$ = 288 K and ACS (blue) T$_{ev}$ = 283 K and T$_{con}$ = 298 K. Volumetric heat energy transferred to the condenser per unit of volume of MOF using (c) methanol and (d) ethanol in MAF-6 with variation of the temperature of the evaporator assuming full desorption.}
    \label{fig:fig_11}
\end{figure}

\begin{figure}[t!]
    \centering
    \includegraphics[width=0.48\textwidth]{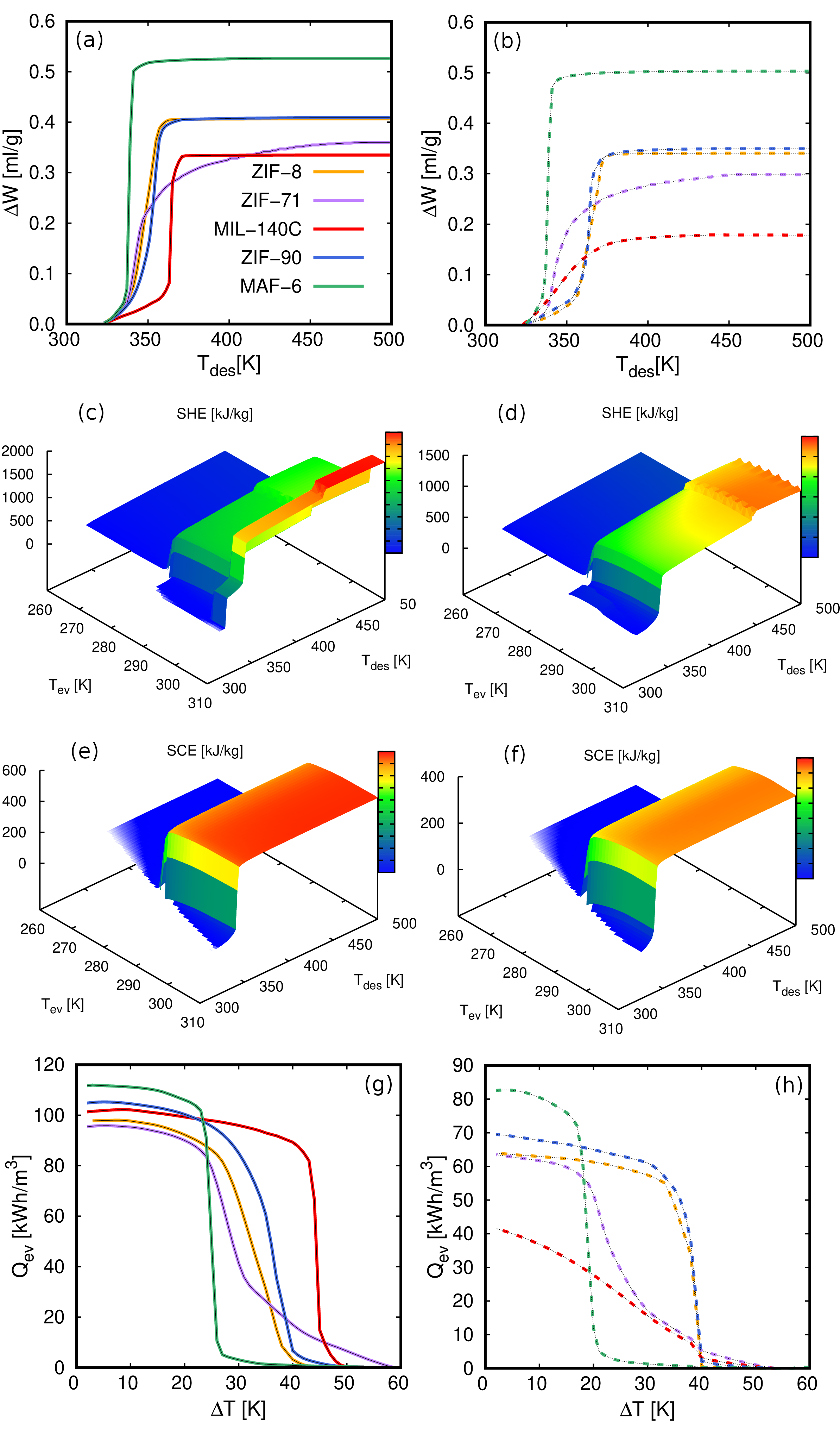}
    \caption{ (a) Deliverable capacity for methanol (solid lines) and ethanol (dashed lines) as a function of the desorption temperature for T$_{ev}$ = 300 K and T$_{con}$ = 310 K, (b) and (c) are SHE and SCE of methanol (top) and ethanol (bottom) in MAF-6 with T$_{con}$ = 313 K, and (d) Q$_{ev}$ as a function of temperature lift (T$_{con}$ - T$_{ev}$), for methanol (solid lines) and ethanol (dashed lines) varying T$_{ev}$ with T$_{con}$ and T$_{des}$ set to 298 K and 373 K, respectively.}
    \label{fig:fig_12}
\end{figure}

In line with the previous results, we can use the multistep approach to analyze other relevant properties for heating or cooling applications as a function of the desired conditions. These properties are, for example, the specific heating (SHE) or cooling effect (SCE), the working volume or deliverable capacity ($\Delta$W), or the heat released or required by the evaporator (Q$_{ev}$). The values for these properties obtained in MAF-6 for methanol and ethanol are shown in Figure 12. The values obtained for the other MOFs can be found in the ESI (Figures S10-S12). As mentioned above, methanol performs better than ethanol in all the properties studied here due to its optimal molecular packing. We also found that all properties are very sensitive to the operating conditions, moving from low to high values in a minimal interval. Because of the number of variables defining the thermodynamic cycle, these drastic changes cannot be predicted from the behavior of the adsorption isotherms or isobars. Under these circumstances, our multistep method that covers all the conditions simultaneously proves to be an efficient technique for analyzing the performance of adsorption-based energy storage devices. For example, the complex landscape of the SHE of methanol and ethanol in MAF-6 (Figures 12b and c) can lead to very different values when setting some of the operational temperatures. This behavior is similar to that in the COP shown in Figure 8. In a simpler comparison, setting working conditions to specific values, MAF-6 would not have appeared as the best performance adsorbent. However, among all MOFs, MAF-6 shows the highest values for all the analyzed quantities. In general, all working pairs studied here show good performance for AHP and ACS applications. However, considering the energy release in the condenser and evaporator, the coefficient of performance, and the highest adsorption and release capacity, the best choice would be the MAF-6.

%%%%%%%%%%%%%%%%%%%%%%%%%%%%%%%%%%%%%%%

\section{Conclusions}
\label{sec:conclusion}

We shed light on the mechanisms governing adsorption-driven heat pumps for heating and cooling applications using MOFs and light alcohols. To this aim, we evaluated the performance of five MOFs using methanol and ethanol as working fluids. Adsorption isotherms and isobars, energetic interactions between molecules and adsorbents, and nucleation of the fluids in confinement were calculated from simulation. Adsorption data was processed using mathematical modeling based on the Dubinin-Polanyi theory of adsorption and a thermodynamic model to describe relevant properties for heating and cooling applications. Finally, we proposed a multistep approach to analyze the relationships between performance and operating conditions, which allows describing the optimal working conditions for each adsorbent-fluid pair.  

All MOFs selected, combined with methanol, show high performance for AHP and ACS applications. The performance coefficients were above 0.8 for cooling and 1.8 for heating. The energy released to the condenser was above 90 kWh/m$^3$. The yield associated with ethanol adsorption is lower than for methanol but still significant for MAF-6. This MOF outperforms the other adsorbents studied here for AHP and ACS applications in a wide range of operating conditions. It exhibits COP above 0.9 and 1.9 for cooling and heating, respectively, and Q$_{con}$ and Q$_{ev}$ about 115 kWh/m$^3$ for methanol. With a pore size of about 18 Å and relatively low density,  the hydrophobic MAF-6 exhibits large pore volume and surface area, resulting in a steeped isotherm and large adsorption capacity for light alcohols. 

Overall, the multistep process proposed here seems to be an efficient tool for analyzing the performance of working pairs for heating and cooling applications. We have proven the importance of removing constraints for applying the thermodynamic model. Establishing or assuming fixed values of certain operating temperatures could lead to misinterpretation of the performance of the working pairs. This is particularly remarkable for systems showing steeped isotherms, which at the same time are desired for these applications. An increase of temperature of a few degrees could decrease by 80\% the performance of a working pair. Another advantage of the proposed approach is that it only needs an adsorption isobar or isotherm as input. This approach can be combined with simulation data or experimental measurements since we employed mathematical modeling to post-process the adsorption data. We show that we can simultaneously describe experimental results from the literature with high accuracy and predict various properties involved in heating and cooling applications.

%%%  END OF MAIN TEXT    %%%%%%%%%%%%%%%%%%%%%%%%%%%%%%%
%%%%%%%%%%%%%%%%%%%%%%%%%%%%%%%%%%%%%%%%%%%%%%%%%%%%%%%%
%%%%%%%%%%%%%%%%%%%%%%%%%%%%%%%%%%%%%%%%%%%%%%%%%%%%%%%%
%%%%%%%%%%%%%%%%%%%%%%%%%%%%%%%%%%%%%%%%%%%%%%%%%%%%%%%%

\section*{Supporting Information}

Electronic Supplementary Information (ESI) available: Structural properties of the adsorbents (Fig. S1 and Table S1), additional details of the models and adsorption conditions (Fig. S2 and Table S2), structural and energetic properties of the adsorbates in confinement (Figs. S3 and S4), comparison between computed and predicted isotherms (Fig. S5), and heat energy transferred to the condenser (Figs. S6-S9), deliverable capacity (Fig. S10), and specific cooling and heating effect of methanol and ethanol (Figs. S11 and S12).

\section*{Acknowledgements}

This work was supported by Ministerio de Ciencia, Innovación y Universidades (CTQ2017-95-173-EXP). A.L-T and S.C. acknowledges funding by the Irène Curie Fellowship program of the Eindhoven University of Technology, and J.M.V-L acknowledges the APSE department of TU/e for funding support and computing resources. We thank C3UPO for the HPC support.

\section*{Author contributions}

R.M. M.-C. and A. L.-T. contributed equally to this work. J.M. V.-L. organized and supervised the project. All authors provided feedback on the interpretation of the results, data analysis, and help writing and revising the manuscript.

\section*{Competing interests}

The authors declare no competing interests.

\bibliography{heat-pumps} % Produces the bibliography via BibTeX.

%Uncomment to add the TOC figure at the end of the manuscript
%\newpage

%\section*{Table of content image}

%\begin{figure}[t!]
%    \centering
%    \includegraphics{Figs/TOC.png}
%\end{figure}

\newpage

\renewcommand{\figurename}{Figure.}
\renewcommand{\thetable}{S\arabic{table}}  
\renewcommand{\thefigure}{S\arabic{figure}} 
\setcounter{figure}{0}

\onecolumngrid 

\begin{center}
    \textbf{\Huge{Supporting Information}}

    \vspace{1.0cm}

    for

    \vspace{1.0cm}

    \textbf{\Large{Alcohol-Based Adsorption Heat Pumps using Hydrophobic Metal-Organic Frameworks}}
\end{center}

    \vspace{2cm}

\begin{figure}[h!]
    \centering
    \includegraphics[width=0.75\textwidth]{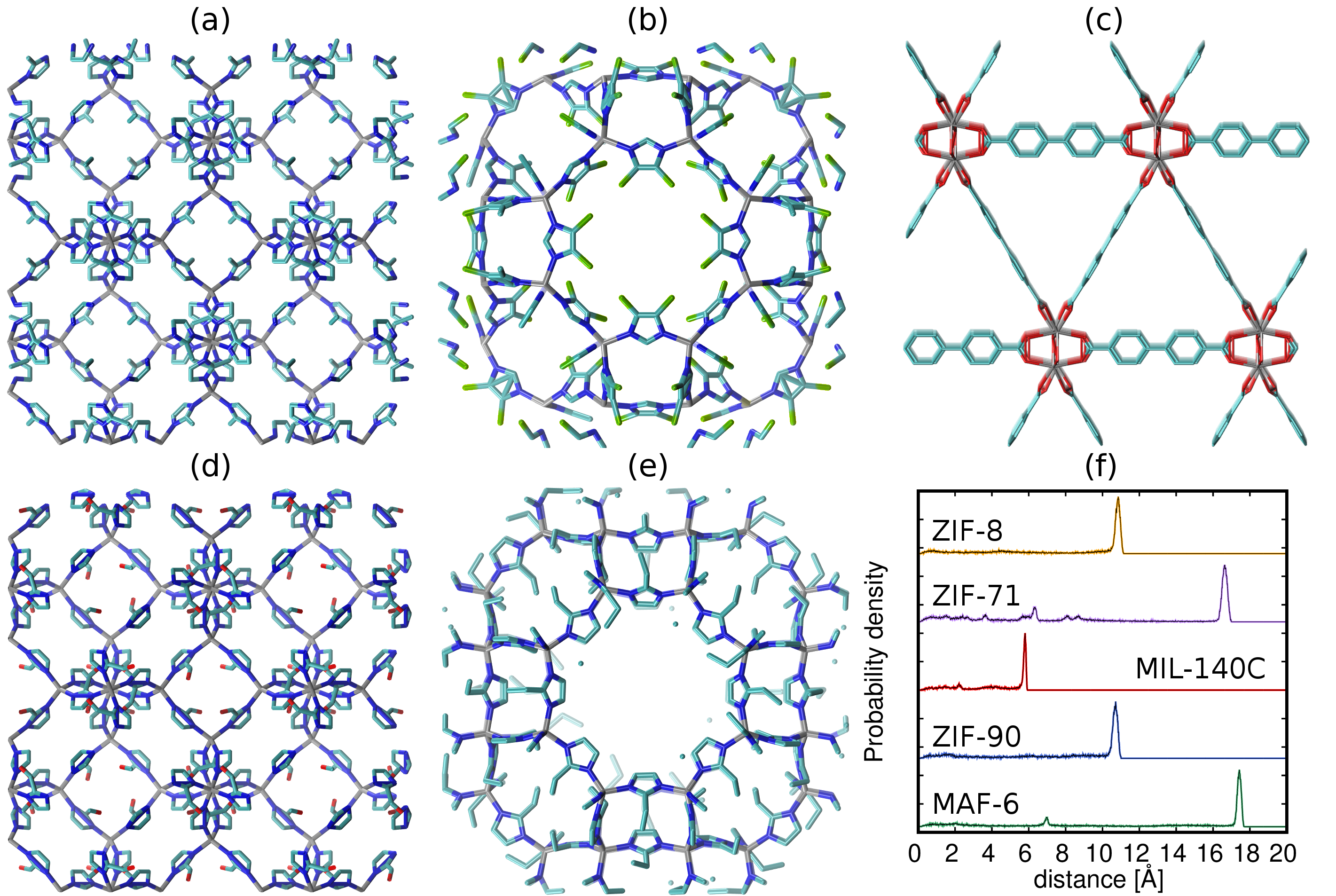}
    \caption{ Schematic representation of the framework connectivity of (a) ZIF-8, (b) ZIF-71, (c) MIL-140C, (d) ZIF-90 and (e) MAF-6. (f) Pore Size Distribution. Nitrogen atoms are coloured in blue, carbon atoms in cyan, oxygen atoms in red, chlorine atoms in green, and zirconium and zinc atoms in grey. Hydrogen atoms are omitted for clarity.}
    \label{fig:fig_S01}
\end{figure}

\noindent Table S1. Structural properties of the selected MOFs, density, pore volume, helium void fraction, surface area, and pore size.

\begin{figure}[h!]
    \centering
    \includegraphics[width=0.75\textwidth]{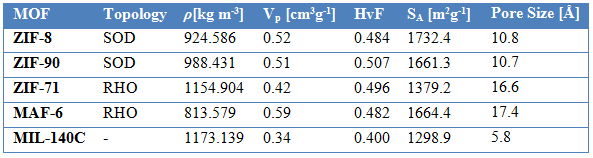}
\end{figure}

\newpage

\begin{figure}[h!]
    \centering
    \includegraphics[width=0.85\textwidth]{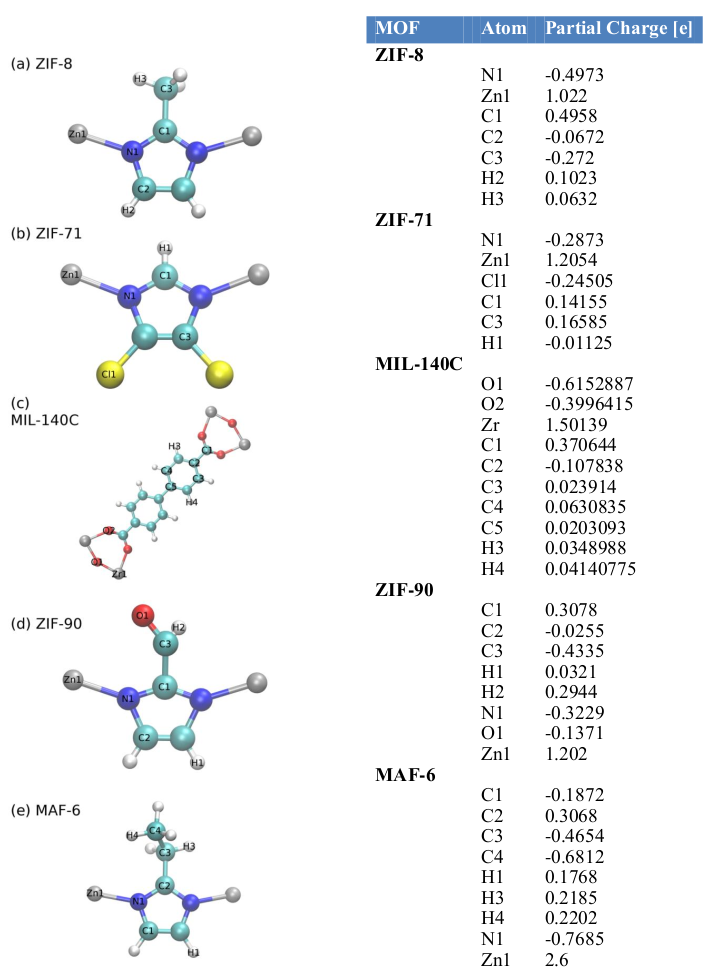}
    \caption{ Schematic representation of the organic linkers of (a) of ZIF-8, (b) ZIF-71, (c) MIL-140C, (d) ZIF-90 and (e) MAF-6. Nitrogen atoms in blue, carbon atoms in cyan, oxygen atoms in red, chlorine in yellow, hydrogen in white and zirconium and zinc atoms in grey. Partial charges of atoms of each MOF.}
    \label{fig:fig_S02}
\end{figure}

\newpage

\newpage

\noindent Table S2. Equilibrium adsorption conditions of methanol and ethanol for the selected MOFs.

\begin{figure}[h!]
    \centering
    \includegraphics[width=0.5\textwidth]{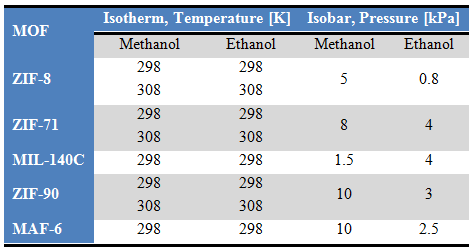}
\end{figure}

\begin{figure}[h!]
    \centering
    \includegraphics[width=0.30\textwidth]{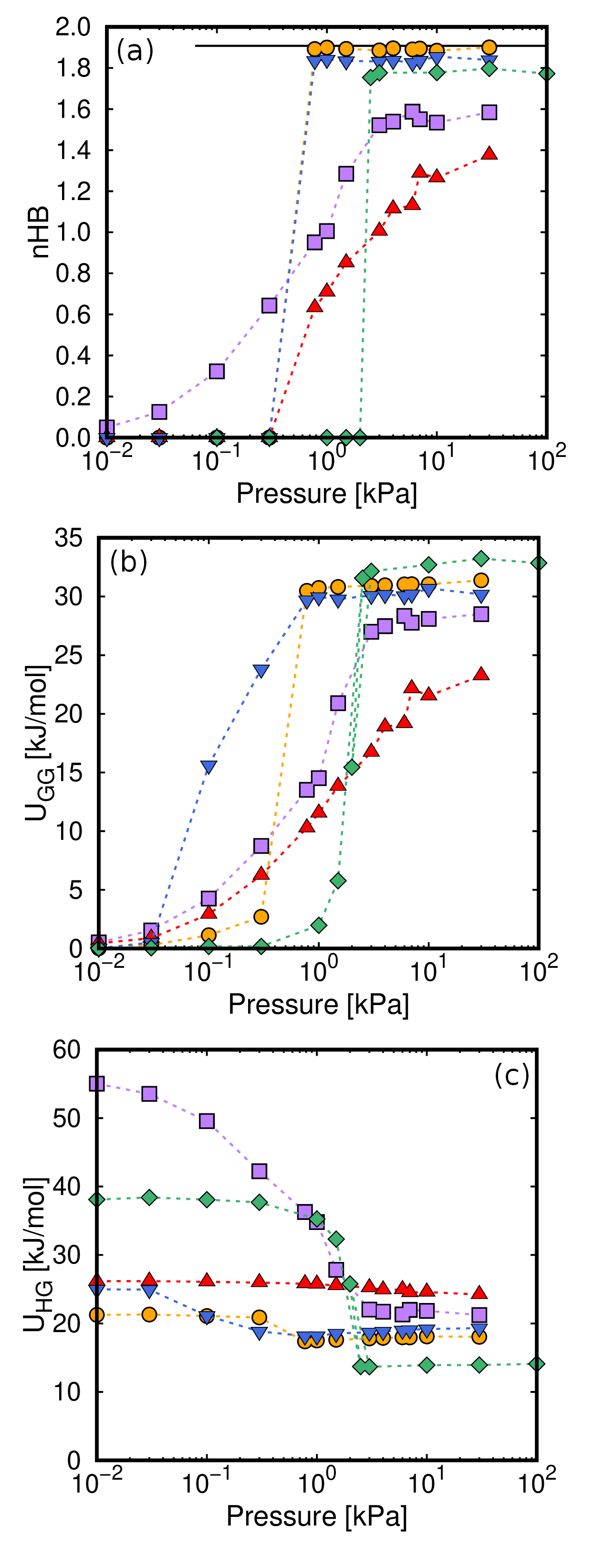}
    \caption{ (a) nHB, (b) guest-guest interactions, and (c) host-guest interactions for ethanol as a function of the pressure. Non-depicted error bars denotes fluctuations smaller than symbol size.}
    \label{fig:fig_S03}
\end{figure}

\begin{figure}[h!]
    \centering
    \includegraphics[width=0.75\textwidth]{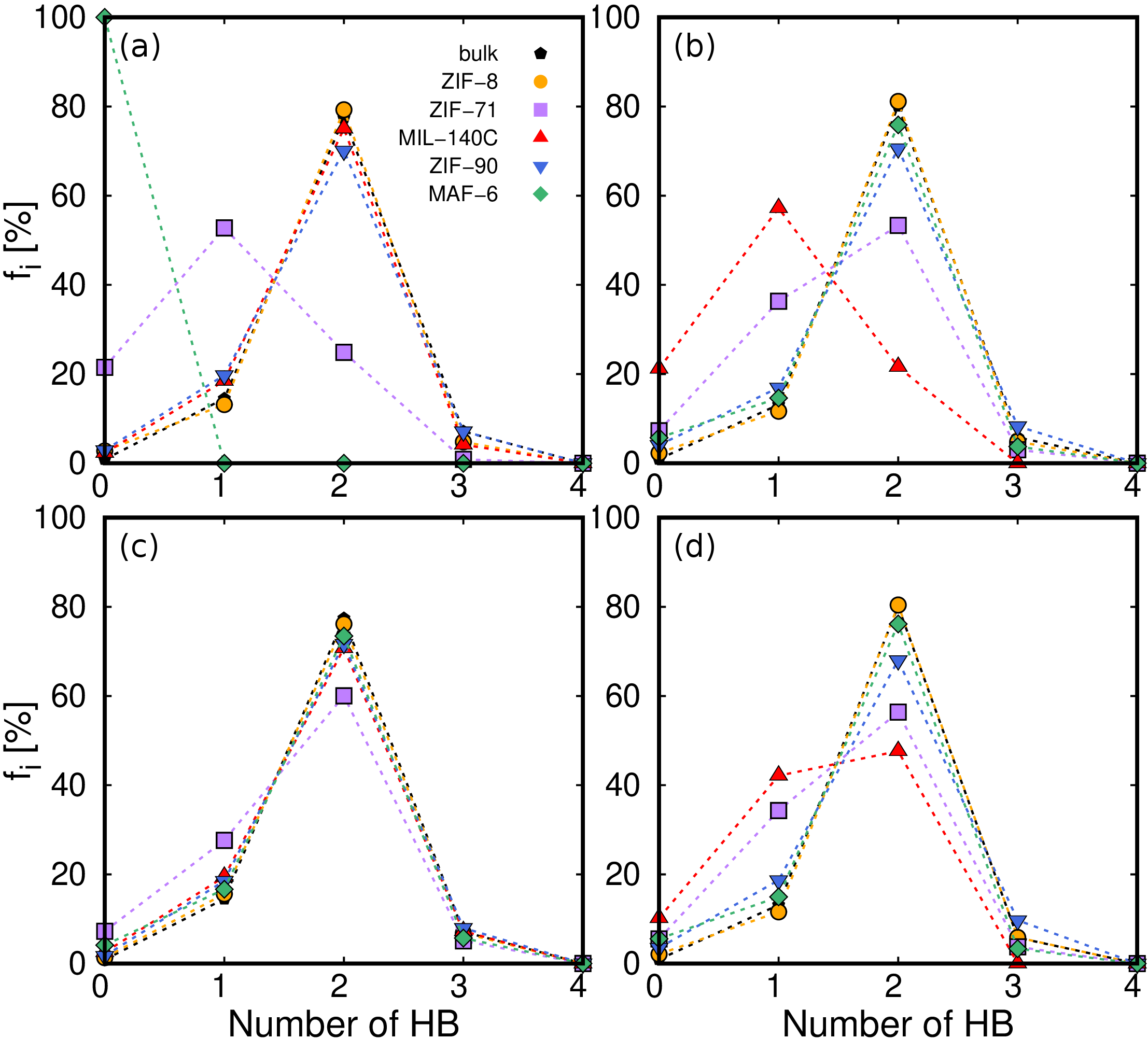}
    \caption{ f$_i$ (percentage of molecules with n HBs) of (a) methanol and (b) ethanol at 3kPa. (c, d) the same at saturation conditions. Non-depicted error bars denotes fluctuations smaller than symbol size.}
    \label{fig:fig_S04}
\end{figure}

\begin{figure}[h!]
    \centering
    \includegraphics[width=0.55\textwidth]{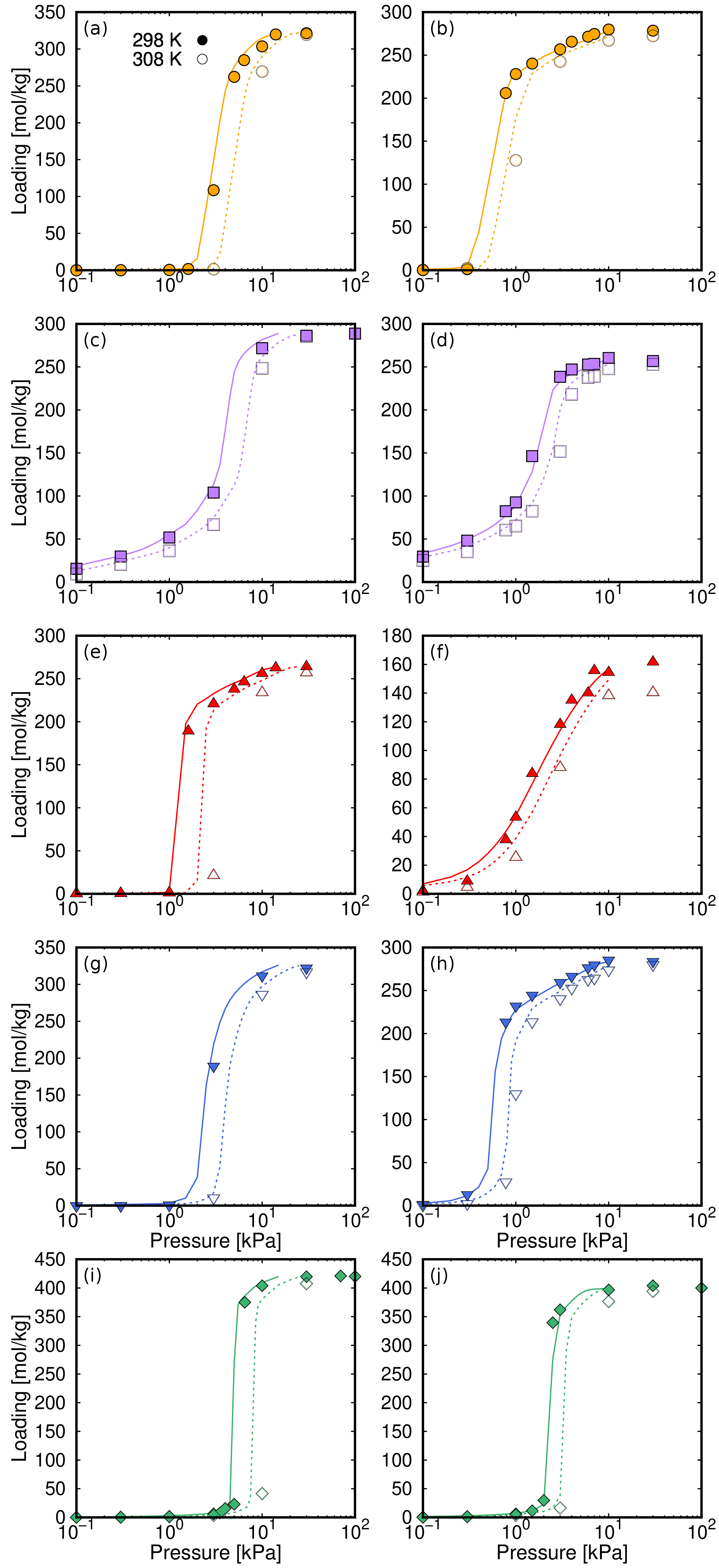}
    \caption{ Adsorption isotherms calculated with GCMC simulations (symbols) and  from the characteristic curve (lines) of methanol (left column) and ethanol (right column) in (a, b) ZIF-8, (c, d) ZIF-71, (e, f) MIL-140C, (g, h) ZIF-90, and (i, j) MAF-6 at 298 K and 308 K.}
    \label{fig:fig_S05}
\end{figure}

\begin{figure}[h!]
    \centering
    \includegraphics[width=0.8\textwidth]{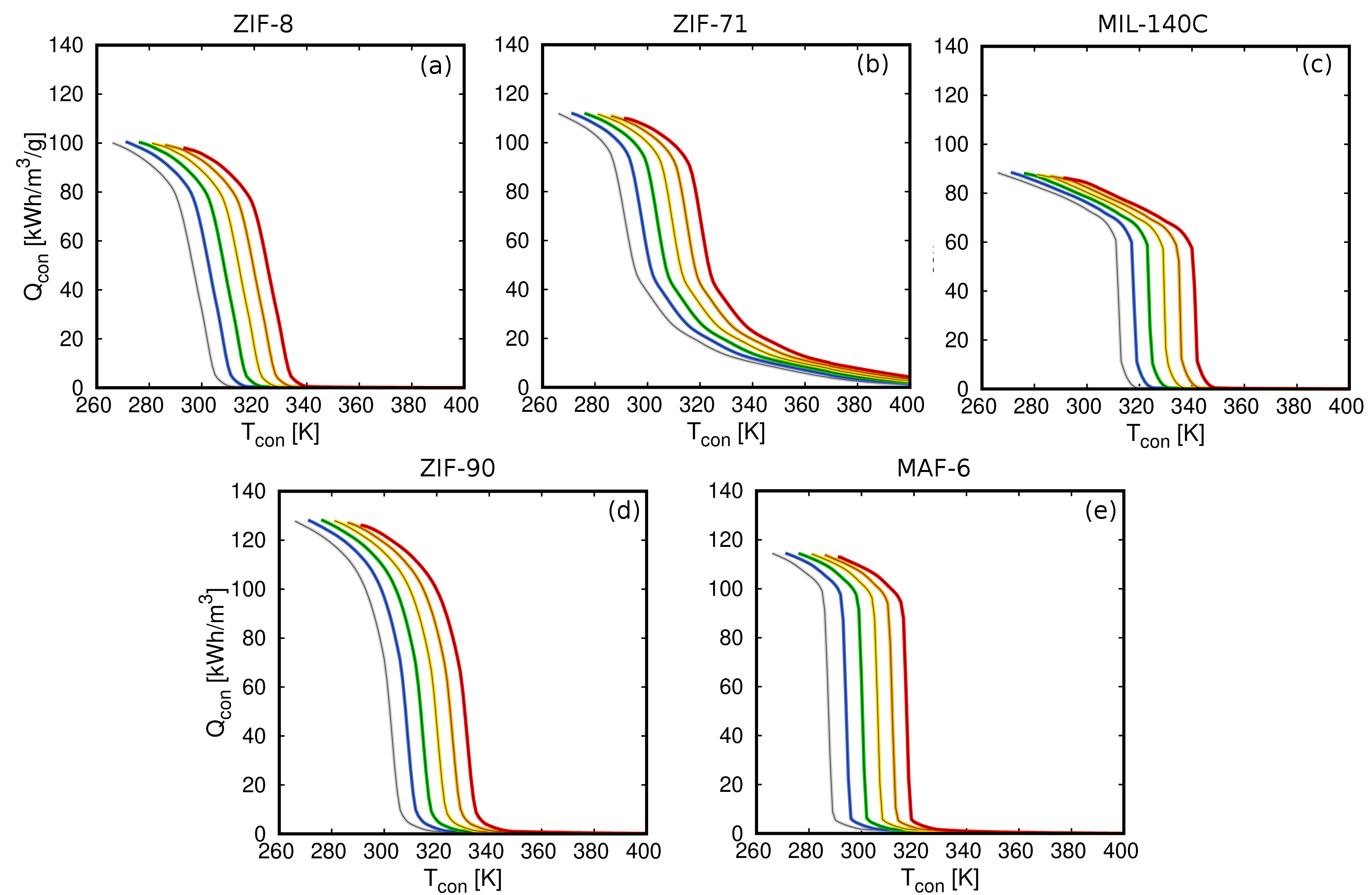}
    \caption{ Volumetric heat energy transferred to the condenser (Q$_{con}$) per unit of volume of MOF using methanol in all MOFs with variation of the temperature of the evaporator assuming full desorption. }
    \label{fig:fig_S06}
\end{figure}

\begin{figure}[h!]
    \centering
    \includegraphics[width=0.8\textwidth]{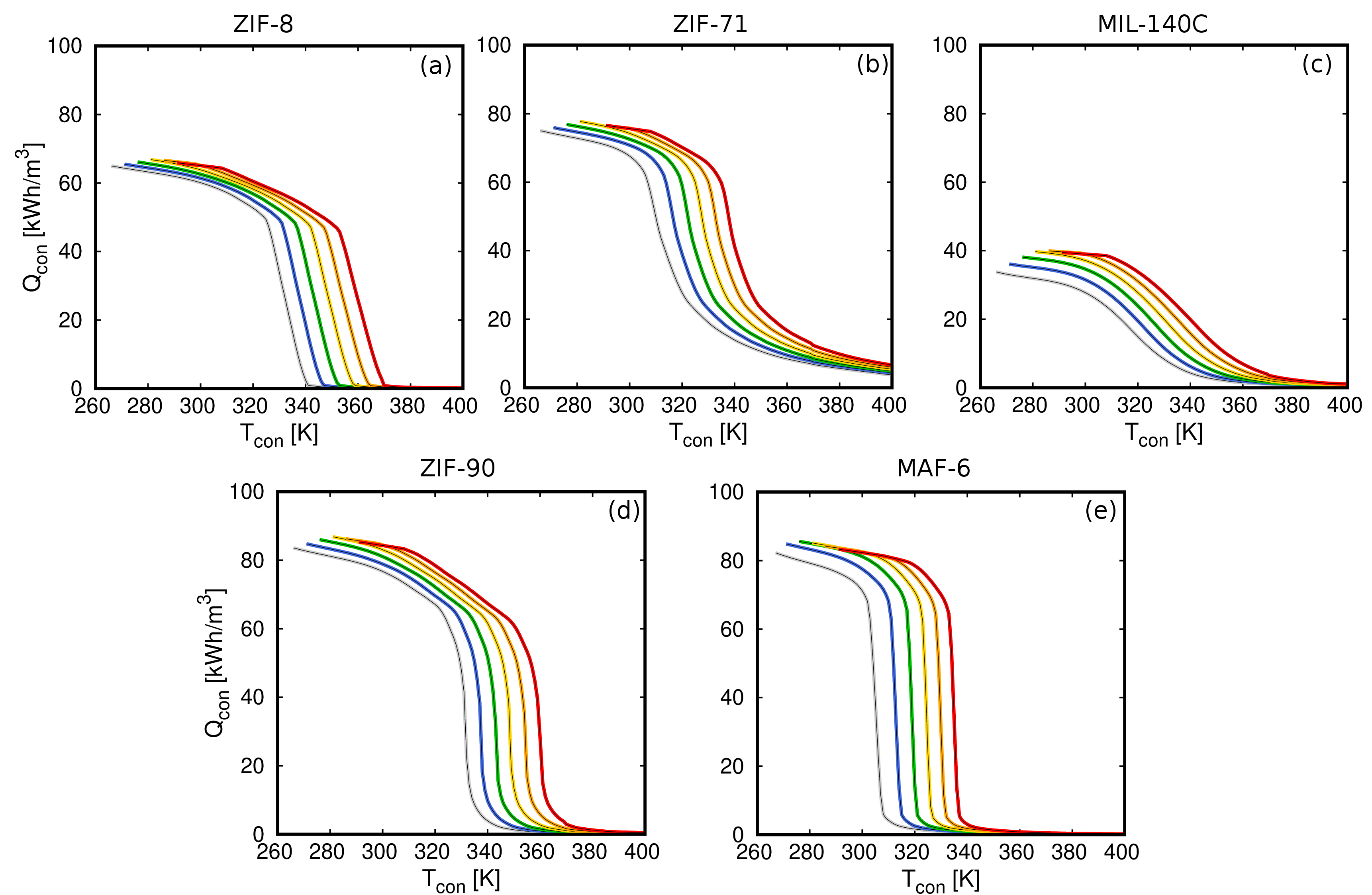}
    \caption{ Volumetric heat energy transferred to the condenser (Q$_{con}$) per unit of volume of MOF using ethanol in all MOFs with variation of the temperature of the evaporator assuming full desorption. }
    \label{fig:fig_S07}
\end{figure}

\begin{figure}[h!]
    \centering
    \includegraphics[width=0.8\textwidth]{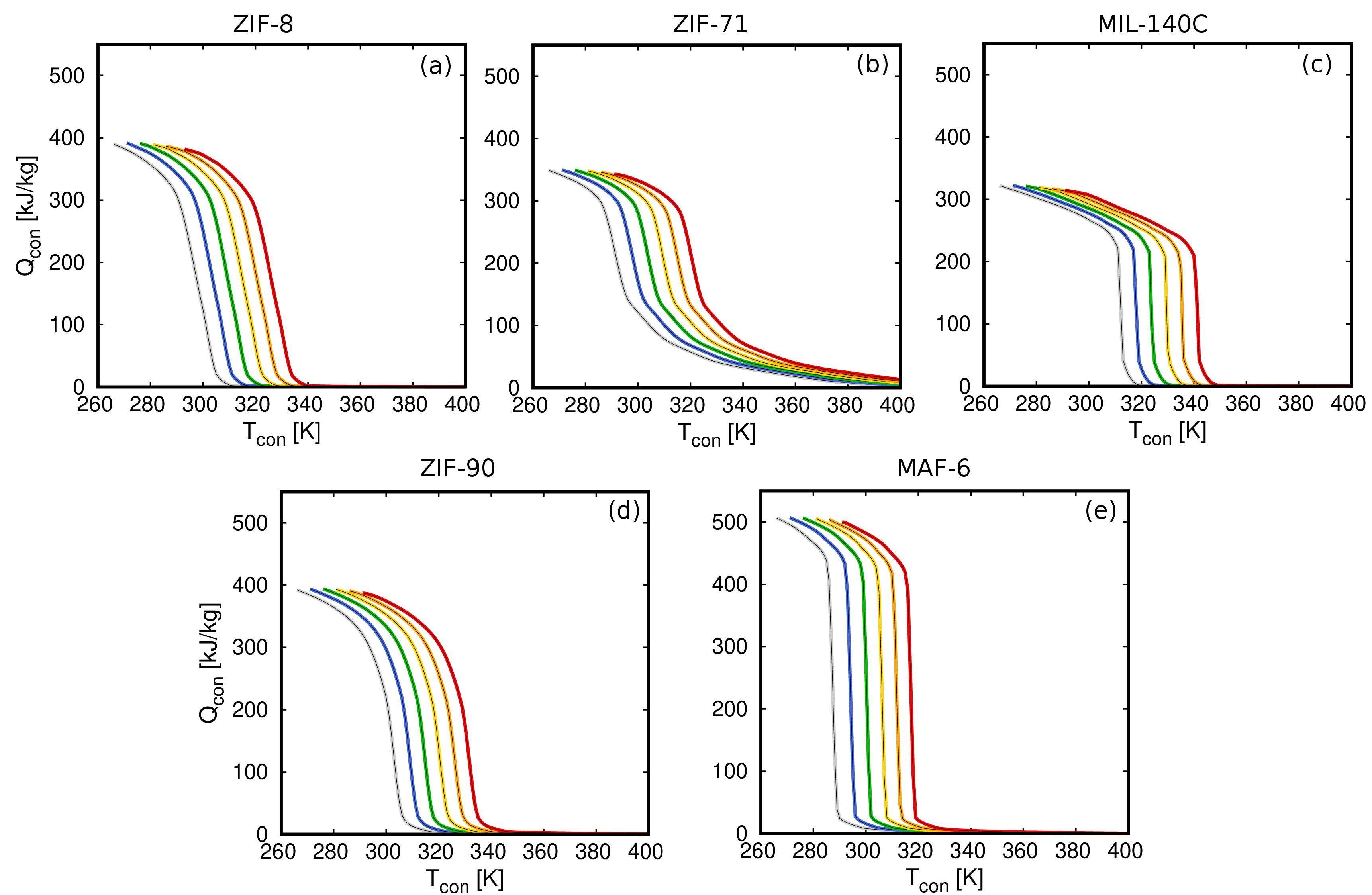}
    \caption{ Gravimetric heat energy transferred to the condenser (Q$_{con}$) per unit of volume of MOF using methanol in all MOFs with variation of the temperature of the evaporator assuming full desorption. }
    \label{fig:fig_S08}
\end{figure}

\begin{figure}[h!]
    \centering
    \includegraphics[width=0.8\textwidth]{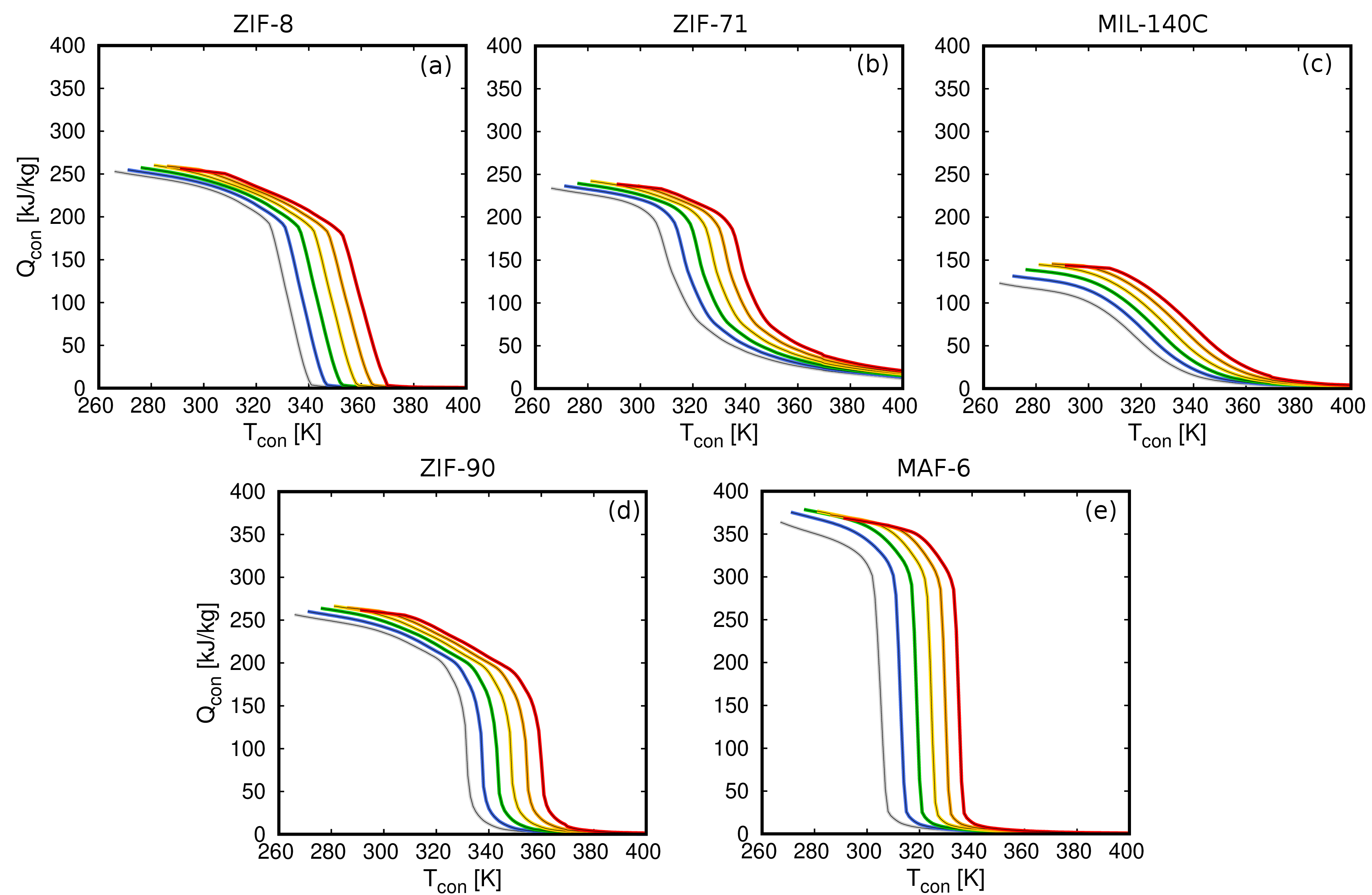}
    \caption{ Gravimetric heat energy transferred to the condenser (Q$_{con}$) per unit of volume of MOF using ethanol in all MOFs with variation of the temperature of the evaporator assuming full desorption. }
    \label{fig:fig_S09}
\end{figure}

\begin{figure}[h!]
    \centering
    \includegraphics[width=0.6\textwidth]{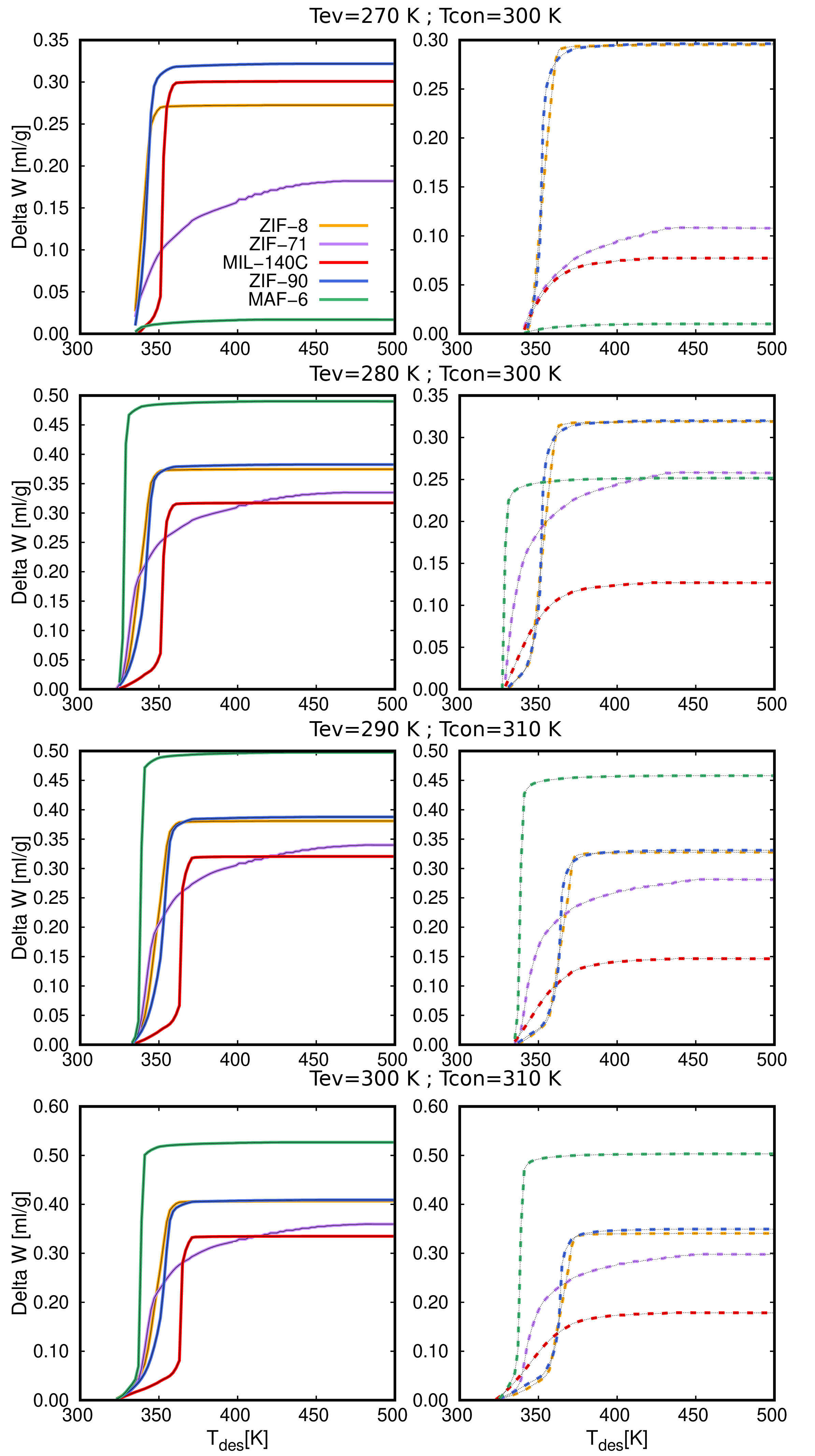}
    \caption{ Deliverable capacity for methanol (solid lines) and ethanol (dashed lines) as a function of the desorption temperature for varying the temperatures of the evaporator and condenser.  }
    \label{fig:fig_S10}
\end{figure}

\begin{figure}[h!]
    \centering
    \includegraphics[width=0.7\textwidth]{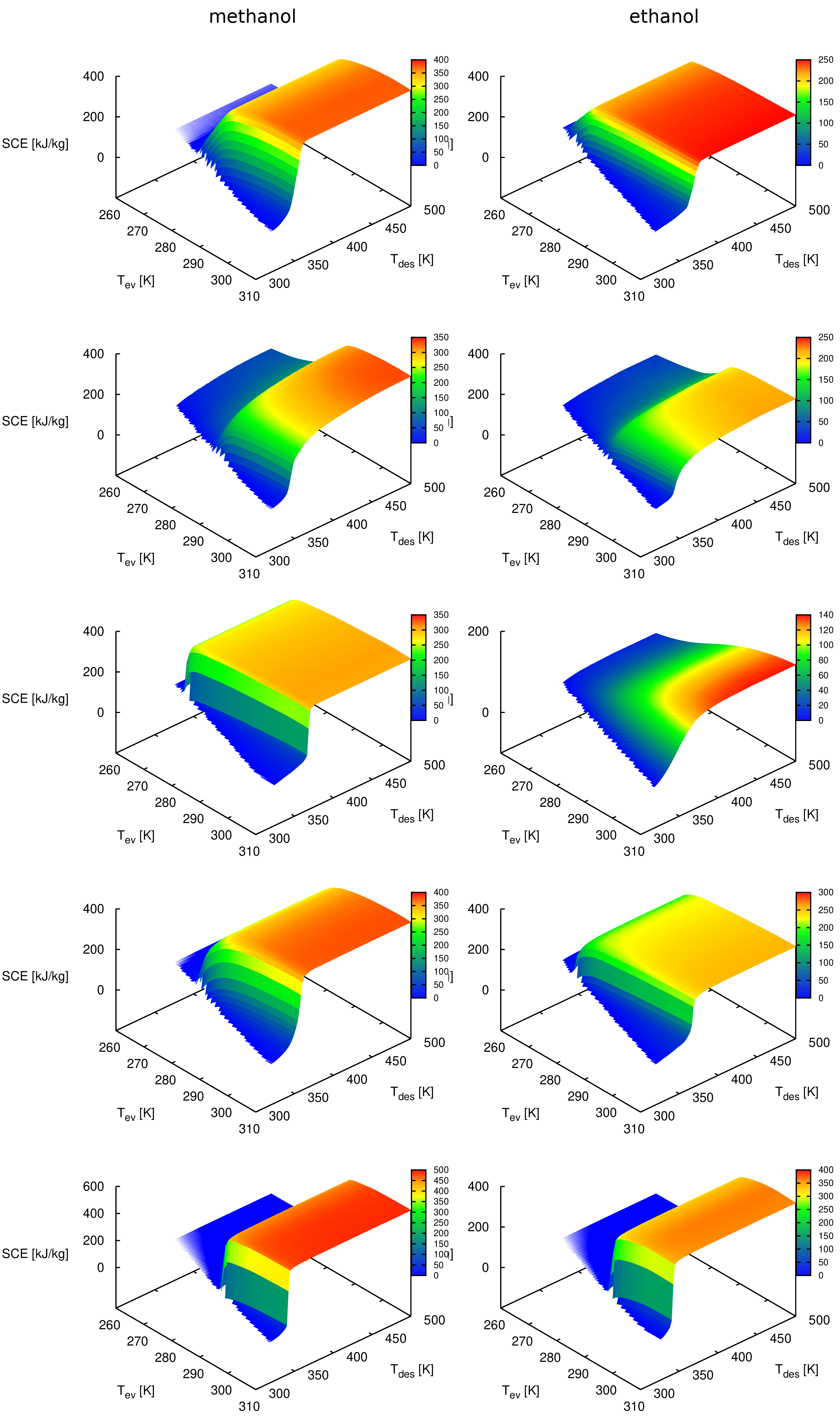}
    \caption{ SCE of methanol and ethanol in all MOFs at T$_{con}$ = 313 K. From top to bottom, ZIF-8, ZIF-71, MIL-140C, ZIF-90, and MAF-6, respectively.  }
    \label{fig:fig_S11}
\end{figure}

\begin{figure}[h!]
    \centering
    \includegraphics[width=0.7\textwidth]{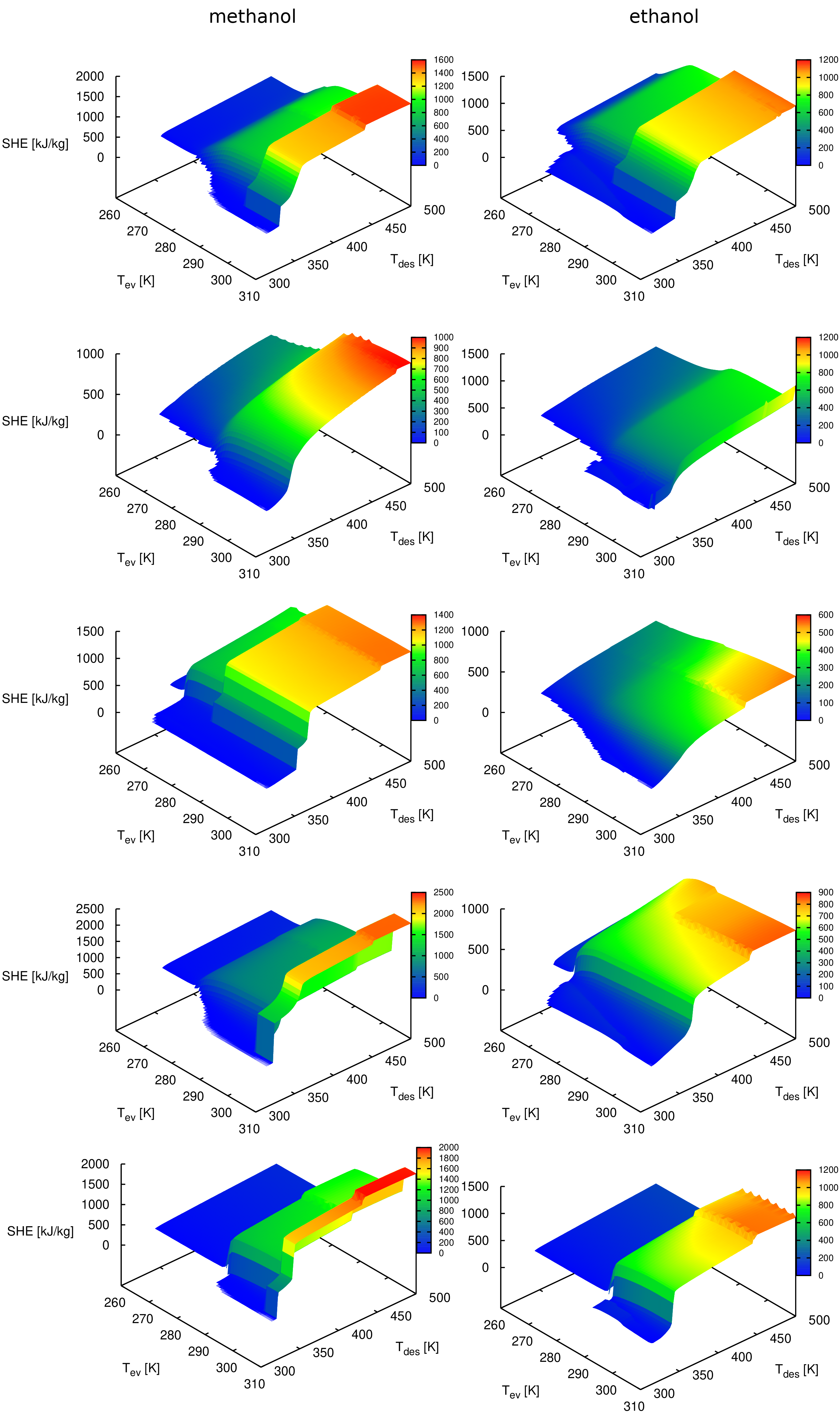}
    \caption{ SHE of methanol and ethanol in all MOFs at T$_{con}$ = 313 K. From top to bottom, ZIF-8, ZIF-71, MIL-140C, ZIF-90, and MAF-6, respectively. }
    \label{fig:fig_S12}
\end{figure}

\end{document}